\documentclass{article}[10pt]
\usepackage[utf8,utf8x]{inputenc}
\usepackage{authblk}
\usepackage{enumitem,amssymb}
\usepackage{graphicx,scalerel}
\usepackage{float}
\usepackage{ucs}
\usepackage{amsmath,amsfonts,amsthm}
\usepackage{mathtools}
\usepackage{commath}
\usepackage{dsfont}
\usepackage{xcolor}
\usepackage{hyperref}
\usepackage{url}
\usepackage{ulem}
\usepackage{pst-node}
\usepackage{tikz-cd} 
\usepackage{listings}
\usepackage{makecell,array,booktabs,multirow}
\usepackage{colortbl}
\usepackage{graphicx}
\usepackage{subcaption}

\usepackage{tikz}
\usetikzlibrary{arrows.meta, positioning}
\usepackage{xcolor}
\usetikzlibrary{trees}

\definecolor{Gray}{gray}{0.85}

\title{Cyclists route choice modeling from trip duration data in urban areas}

\date{}

\author[1]{Bertrand Jouve}
\author[2]{Paul Rochet} 
\author[1,2]{Mohamadou Salifou}

\affil[1]{LISST, Université de Toulouse, Université de Toulouse 2, CNRS,  Toulouse, France.}
\affil[2]{F\'ed\'eration ENAC ISAE-SUPAERO ONERA, Universit\'e de Toulouse, France.}

\begin{document}

\maketitle

\begin{abstract}
The lack of GPS data limits the ability to reconstruct the actual routes taken by cyclists in urban areas. This article introduces an inference method based solely on trip durations and origin–destination pairs from bike-sharing system (BSS) users. Travel time distributions are modeled using  log-normal mixture models, allowing us to identify the presence of distinct behaviors. The approach is applied to 3.8 million trips recorded in 2022 in the Toulouse metropolitan area, with observed durations compared against travel times estimated by OpenStreetMap (OSM). Results show that, for many station pairs, trip durations align closely with the fastest route suggested by OSM, reflecting a dominant and routine practice. In other cases, mixture models reveal more heterogeneous behaviors, including longer trips, detours, or intermediate stops. This approach highlights both the stability and diversity of cycling practices, providing a robust tool for usage analysis in data-limited contexts, and offering new insights into urban mobility dynamics without relying on spatially explicit data.

\end{abstract}








\section{Introduction}

Cycling is experiencing a growing momentum, both in countries where it has not historically been a major mode of transport and in those where its use is already well established \cite{harms2018cycling, majumdar2019study}. Importantly, a trip in bicycle is not merely a means of getting from point A to point B, but has an intrinsic value in itself, reflecting a fundamental human need for mobility \cite{mokhtarian2001understanding}. This development is part of public policies that acknowledge the positive externalities of cycling, whether environmental or health-related, while also contributing to improved urban quality of life and reduced road congestion. Moreover, several studies have shown that the subjective experience of cycling tends to generate greater satisfaction compared to other modes of transport \cite{de2018towards,harms2018cycling,yan2024cycling, mohamed2019speed}. \\

Understanding cyclists’ route choice behavior remains a key issue for planning appropriate cycling infrastructure. Route choice is a complex decision-making process in which individuals select a specific path between an origin and a destination based on multiple simultaneous attributes, such as efficiency, enjoyment, perceived safety, or the presence of dedicated facilities \cite{sener2009analysis,hood2011gps,broach2012cyclists,casello2014modeling,  wuerzer2015cycling,bapat2017route,chen2018gps,zimmermann2017bike}. To model these choices, 
stated preference surveys or data collected from Global Positioning Systems (GPS) are typically used. While surveys make it possible to collect information directly from users, their accuracy remains limited as they rely on participants' memory and perception of environment \cite{winters2010far,kerr2011using,krenn2014route,sun2017built,schantz2017distance}. In contrast, GPS data allow precise real-time tracking of the routes taken, with detailed insights into distances, speeds, and durations \cite{lu2018understanding,kim2021analysis,scott2021route}. Furthermore, when combined with Geographic Information Systems (GIS), they enable the spatial contextualization of observed behaviors \cite{geurs2004accessibility,chung2024cycling}. However, such data are rarely available in bike-sharing systems (BSS), which generally lack embedded tracking sensors. This creates a major methodological challenge: how can cyclists’ behaviors be inferred from trip durations in the absence of route traces? \\

This is the question that guides our investigation, focusing on trips made with the Toulouse bike-sharing system (BSS). The proposed approach relies on an original inference strategy: using only trip durations and origin–destination pairs, we aim to identify regularities in user behavior. To this end, we compare observed durations with the estimated times provided by the OpenStreetMap (OSM) platform, which generates optimized routes according to different criteria (notably the shortest and the fastest). These comparisons are combined with log-normal models and, when necessary, log-normal mixture models, in order to better account for the heterogeneity of observed durations and to detect the existence of differentiated practices.\\

This approach contributes to an active field of research on detours in cycling mobility, which has long focused on distance-based analyses. Numerous studies have compared actual routes with the shortest-distance paths \cite{kerr2011using,broach2012cyclists, kelly2013assessing,huang2015circuity,rupi2019data,park2019bicyclists,dehaas2020cycling,costa2021circuity,chou2023analysis,chung2024cycling}. Yet, to our knowledge, very few empirical studies have explored this issue through the lens of travel time even though travel duration is often the only available metric in most BSS datasets \cite{mythesissmit2021, report2008}. This gap is especially problematic given that travel time plays a central role in mobility decisions. \cite{sener2009analysis} have shown that travel time (as well as heavy traffic volume) are the most important attributes in bicycle route choice. In their report, \cite{report2008} comes to the same conclusion and goes on to say that the data from their study showed that the bicycle is competitive in terms of travel time with the automobile for many short trips. \cite{broach2012cyclists} highlighted a slight difference between the distance traveled and the actual travel time. This was confirmed by \cite{Hull01012014} who categorises both the shortest and fastest route under route directness. Although these two types of routes generally overlap, several factors can negatively affect travel time, leading cyclists to deviate from the shortest path.\\

Minimizing travel time can be defined as cyclists’ tendency to favor routes that reduce travel duration by avoiding features that cause delays \cite{broach2012cyclists}. This involves not just seeking the shortest route in terms of distance, but more importantly, the one that optimizes time by taking potential obstacles into account. This optimization can be assessed by examining the extent of overlap between the actual route taken and the theoretically fastest route, while also analyzing how much certain time-delaying features influence route choice. Among the elements that can increase travel time are traffic lights, stop signs, and gradients (elevations). \cite{menghini2010route} showed that these factors often lead cyclists to avoid the shortest paths, opting instead for longer routes in distance that are smoother and quicker in practice.\\

Recent literature distinguishes between two main types of detours, namely \textit{realized detours} resulting from network constraints (e.g., lack of bike lanes, physical obstacles, network discontinuities) and \textit{behavioral detours} reflecting a conscious preference for routes perceived as more pleasant or safer, even if they are longer \cite{chou2023analysis}. These detours reveal behavioral logics that traditional transport models often fail to capture, especially when based on the assumption that individuals systematically seek to minimize distance or time. \\

The objective of this research is to better understand the behavior of bike-sharing users through the analysis of observed trip durations. By studying the distributions of travel times, we aim to identify collective trends in cycling practices, such as the existence of a dominant behavior for certain station pairs, as well as situations where no single typical profile emerges. These latter cases may reveal the coexistence of heterogeneous practices, including longer trips, detours, intermediate stops, or alternative route choices. To capture these dynamics, trip duration distributions are modeled using a log-normal distribution \cite{kou2019understanding} or, when necessary, a mixture of log-normal distributions. A goodness-of-fit test ($\chi^{2}$ test) is applied to assess whether a single log-normal distribution adequately represents the data. When this model is rejected, the presence of multiple components suggests distinct groups of trips, potentially corresponding to different routing strategies. Finally, these empirical observations are compared to routes suggested by OpenStreetMap (OSM), providing a meaningful benchmark between actual user practices and theoretical itineraries. This approach allows us to detect potential bicycle detours in the urban environment and their importance, without the need for spatialized data. \\

In the remainder of the paper, Section \ref{methodo} presents the methodological framework, including data preparation from the bike-sharing system, the fitting of log-normal and log-normal mixture models to characterize the variability of trip durations, and the comparison with reference itineraries provided by OpenStreetMap. Section \ref{resul} reports the main results, distinguishing cases of dominant behaviors from those marked by greater heterogeneity, and quantifying the share of users following each type of practice. Section \ref{sec:particular} offers a detailed analysis of selected representative station pairs, illustrating the diversity of observed configurations. Finally, Section \ref{conclu} concludes with a discussion of the contributions and limitations of the approach, as well as its perspectives for research and cycling infrastructure planning. 

\section{Methodology}\label{methodo}

\subsection{Data description and preprocessing}

This study uses data from the bike-sharing system (BSS) implemented in 2007 in Toulouse by JCDecaux. We have access to all Origin/Destination trips completed in 2022. Each record (representing a trip or journey) includes the departure and arrival times (to the nearest minute), the departure and destination stations, along with their geographic coordinates. The dataset also contains information on bike positioning, as well as lock and unlock times, excluding rebalancing operations. As the data are anonymized, it is not possible to track specific users in their daily use of the system. \\

We exclude trips where the origin and destination stations are the same, as well as those lasting 1 minute or less, which accounted for $3.75\%$ of the total data points.
Outliers are defined as values falling outside the interval:
$$
[Q_1 - 1.5 \times \text{IQR},\; Q_3 + 1.5 \times \text{IQR}]
$$
where $Q_1$ and $Q_3$ are the first and third quartiles, and $\text{IQR} = Q_3 - Q_1 $. Observations falling outside this range are excluded from the analysis.
The processed data end up including 10 901 distinct origin/destination station pairs (distinguishing both directions, e.g., (A, B) $\neq $ (B, A)), and a total of 3.8 million trips recorded in 2022 between the 289 dock-stations.

\subsection{Log-normal model on trip durations for a single route} 

Trip durations,truncated to the minute, are observed for all trips between any pair of BSS stations for the year 2022. For a given itinerary, the distribution of the trip durations is extremely well fitted by a log-normal model despite the discrete nature of the data, rounded to the nearest integer. \cite{kou2019understanding} came to the same conclusion, showing that the trip duration follows a log-normal distribution in large bike-sharing systems (e.g. in Boston, Washington DC, Chicago and New York). Surprisingly, these data are less suited to common discrete models such as Negative binomial or Conway–Maxwell–Poisson, for instance.

\begin{figure}[H]
\begin{center}
\includegraphics[width = 0.8\textwidth]{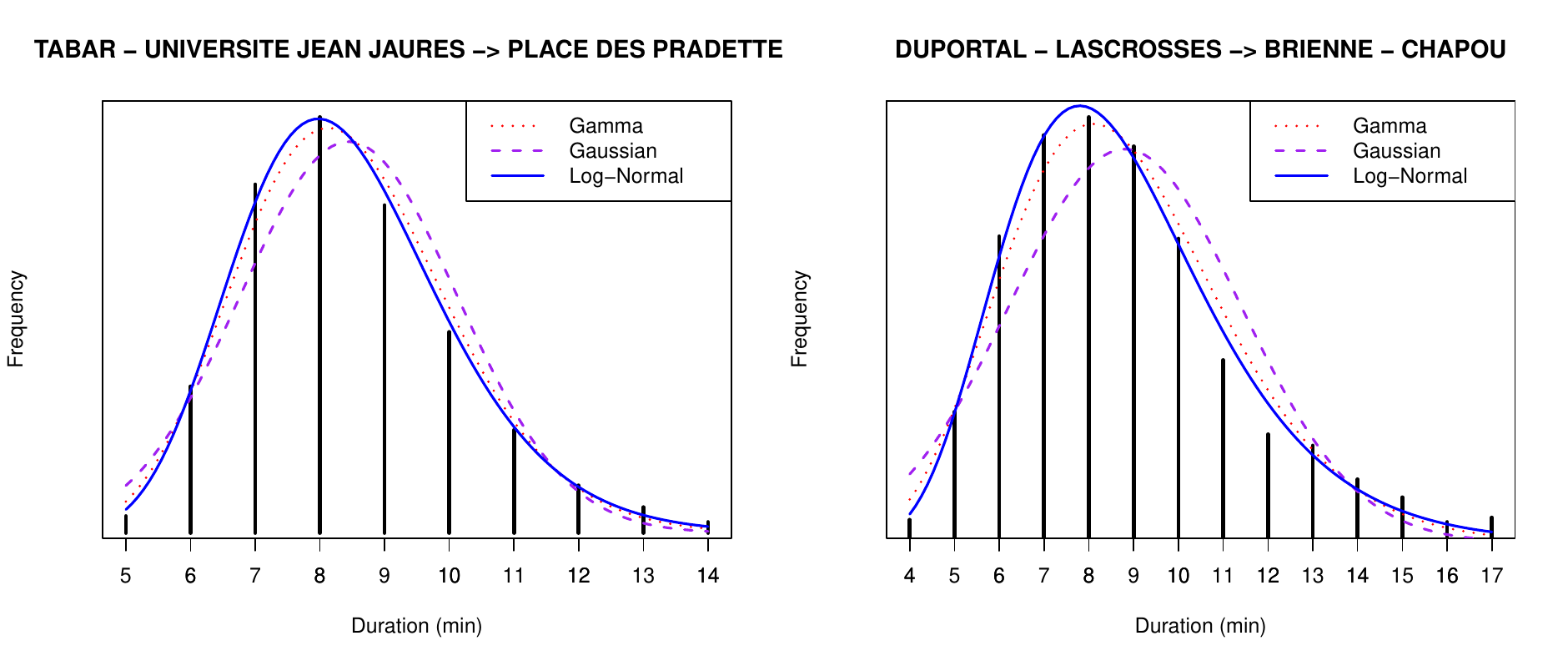}
\end{center}
\caption{\footnotesize Barplots of the trip times (in minutes) from Tabar - Université Jean Jaurès to Place des Pradettes (left) and Duportal Lascrosses to Brienne Chapou (right), with the three fitted models. Real data are plotted with black vertical sticks. Each of the three curves represent an approximation model.}
\label{fig:distri_simple}
\end{figure}
In the two examples of Figure \ref{fig:distri_simple}, data appear to be better fitted by a single log-normal model, compared to Gaussian and Gamma models. This is confirmed by the chi-squared goodness-of-fit test with p-values of $0.07$ and $0.36$ respectively for the log-normal fit, while the p-values are less than $10^{-4}$ for Gaussian and Gamma models. In these cases, there is no practical alternative itinerary between the two stations considered. This suggests the existence of a dominant collective behavior, where most users follow a similar travel pattern reflected in a unimodal distribution of trip times.

\subsection{Test for the coexistence of heterogeneous cycling practices and mixture model}

For a given trip from station $A$ to station $B$, the distribution of observed travel times often displays a degree of heterogeneity that cannot be adequately captured by a single log-normal distribution. This heterogeneity may reflect the coexistence of multiple mobility practices, such as the use of alternative routes, detours, intermediate stops, or time variations caused by different traffic conditions. To formally assess this, we first test whether the empirical distribution of trip durations (truncated to the nearest  minute) can be reasonably approximated by a single log-normal distribution law using a chi-square goodness-of-fit test. Rejection of the null hypothesis is taken as evidence of heterogeneous travel behaviors, prompting us to fit a log-normal mixture model. In our context, a model with two or more components suggests the presence of at least two distinct travel behaviors or routes between the same pair of stations. 

\begin{figure} [H]
    \centering
    \begin{subfigure}{0.45\textwidth}
        \centering
        \includegraphics[width=\textwidth]{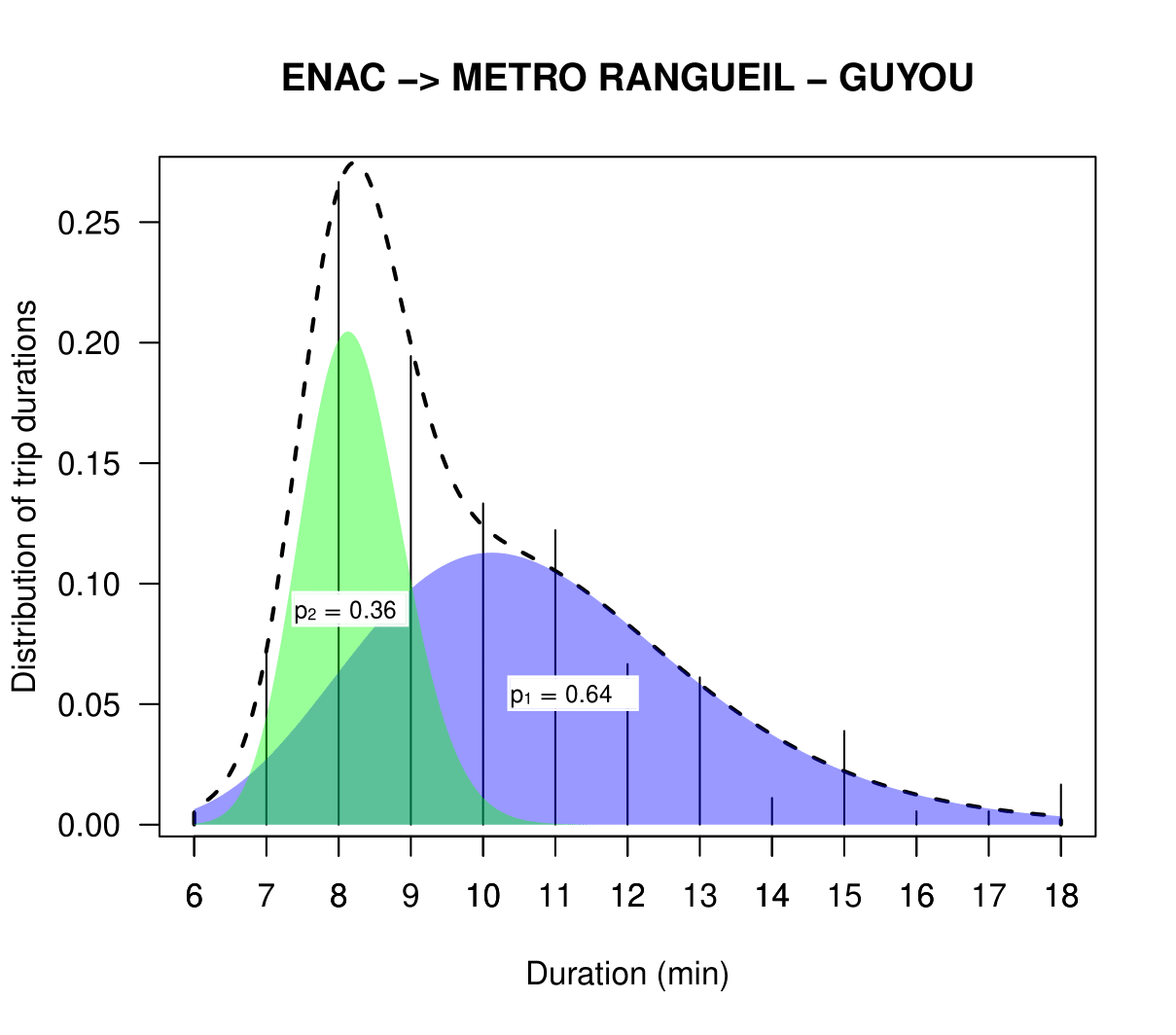}
    \end{subfigure}
    \hfill
    \begin{subfigure}{0.45\textwidth}
        \centering
        \includegraphics[width=\textwidth]{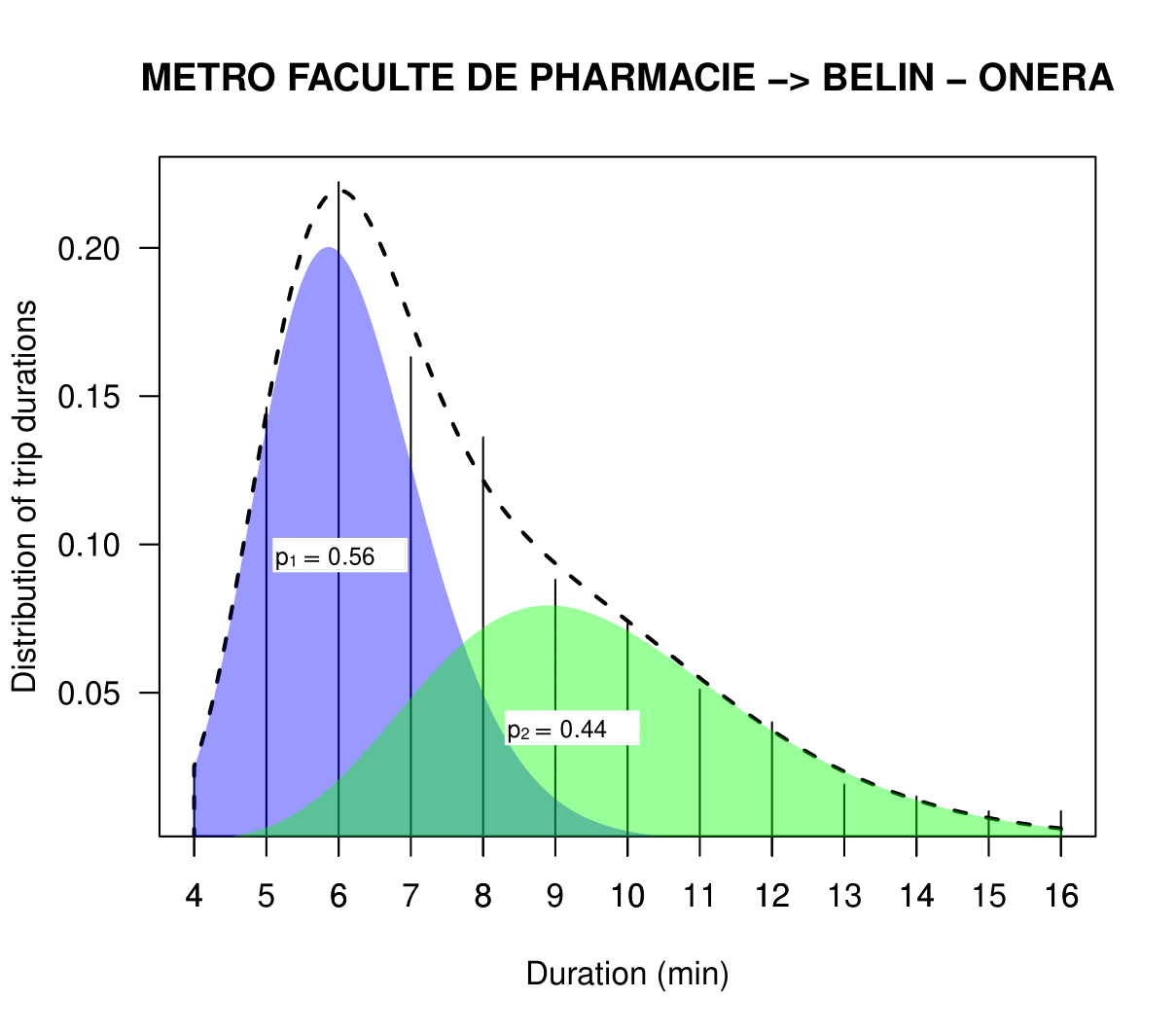}
    \end{subfigure}
        \caption{Log-normal mixture models in presence of alternative routes.}
    \label{fig:lognormal_mix}
\end{figure}

In the two examples of Figure \ref{fig:lognormal_mix}, the distributions suggest that cyclists may adopt at least two different ways of riding between the stations considered, which is reflected by the two components of the log-normal mixture model. In these cases, the null hypothesis of a single log-normal distribution is clearly rejected, while the two-component mixture provides a more faithful representation of the observed variability. \\

For reasons of parsimony and interpretability, we limit the analysis to a maximum of two components in the mixture model, even though the second may actually bring together several distinct practices. In this framework, the first component can generally be interpreted as representing a more direct or time efficient behavior, often associated with the shortest, fastest, or most straightforward route. By contrast, the second component should be understood more broadly, as capturing a range of alternative or less regular cycling practices such as longer trips, detours, intermediate stops, or slower riding conditions rather than as a distinct and homogeneous group. This methodological choice reflects a deliberate compromise: it allows us to account for the disparity of travel times without overfitting, while acknowledging that the components do not necessarily correspond to homogeneous routes but instead to differentiated forms of cycling behavior.

\subsection{Comparison with Open Street Map proposed itineraries}

OpenStreetMap (OSM) is a free, collaborative mapping project launched in 2004. This mapping network is structured according to an arc/node topology, to which a “relationship” attribute is added, allowing interactions between nodes and existing paths to be modeled \cite{mocnik2017openstreetmap}. Data from OpenStreetMap have been integrated into various scientific studies \cite{ibisch2016global} and benefit from continuous updating, carried out on a voluntary basis by a community of contributors. This collaborative dynamic ensures that the cartographic information available is rich and up-to-date. 

This platform offers, through its high-performance routing engine Open Source Routing Machine (OSRM), a range of cyclist profiles (cycling-regular, cycling-road, cycling-mountain, cycling-electric) and preferences (fastest and shortest). To make inferences about specific routes, we compare our data with the travel times of the two main routes provided by OSM, namely:
\begin{itemize}
    \item The \textit{fastest} route, that minimizes the trip duration at the cost of a possibly longer distance. 
    \item The \textit{shortest} route, that minimizes the distance. 
\end{itemize}

To determine the distance (or duration), OSRM uses OSM data to calculate the optimal distance between a departure point and an arrival point based on the selected profile type. The algorithm then finds the path that follows the roads or trails listed in OSM, while respecting the defined constraints (roads forbidden to bicycles, preferences for bike lanes, etc.). The distance calculated in this way corresponds to the physical length of the road or path segments used for the optimized route.
Depending on the selected preference, if the \textit{fastest} option is chosen, OSRM calculates the quickest route by considering average speeds per segment. But, if \textit{shortest} is selected, it favors the shortest possible distance, without taking speeds into account.

Unlike services such as Google Maps, which may incorporate traffic data or user travel history, OSRM relies solely on the geographic and attribute data of roads in OSM. Therefore, the distance calculated by OSRM is theoretical and optimized for the given parameters, and not an average or synthesis of real trips made by cyclists or drivers.
In addition, the speeds used depend on the type of road:
\begin{itemize}
    \item Dedicated cycleways: ~15–20 km/h
    \item Secondary roads with bike lanes: ~15 km/h
    \item Urban roads without dedicated cycling infrastructure: ~12–15 km/h
    \item Dirt roads or trails: ~8–12 km/h
\end{itemize}
	
The estimated durations assume ideal conditions (no stops, no traffic). The estimates may be inaccurate if the data for a region is incomplete or outdated.\\

To complement the statistical analysis, we compare the observed results with the reference routes suggested by OSM. The purpose of this comparison is not to validate a precise route but rather to examine the convergences and divergences between observed behaviors and theoretical routes. This helps illustrate the extent to which dominant practices correspond or not to the optimized paths suggested by OSM, thereby providing insights into the diversity of cycling behaviors in the Toulouse network.

\section{Results} \label{resul}

The analysis was carried out on the entire Toulouse BSS dataset, which comprises 289 stations. Only samples with a minimum of $100$ observations were retained, resulting in $10,901$ directed station pairs. For each of these samples, we tested whether the distribution of travel times could be represented by a single log-normal law using a chi-squared goodness-of-fit test (rounded to the nearest minute). At the significance level $\alpha = 0.05$, a single dominant behavior was identified for $38.18\%$ of station pairs (4163 cases). In most of these situations ($89.77\%$), OSM proposes the same option for both the fastest and the shortest routes, which is consistent with the predominance of one main practice among users.

In contrast, $65.41\%$ of cases revealed evidence of heterogeneous practices through the log-normal mixture model, even though OSM did not differentiate between the fastest and the shortest itineraries. One such example is discussed in Section \ref{sec:particular}. It is worth mentioning that, as an alternative to the chi-squared test, using the Bayesian Information Criterion (BIC) to determine whether one or more components are present in the mixture model leads to similar conclusions in over $96\%$ of cases. 

\subsection{Consistency and diversity in BSS travel behaviors}

\paragraph{Single dominant behavior}~~\\

\begin{figure}[H] 
\begin{center}
\includegraphics[scale = 0.7, width = 0.45\textwidth]{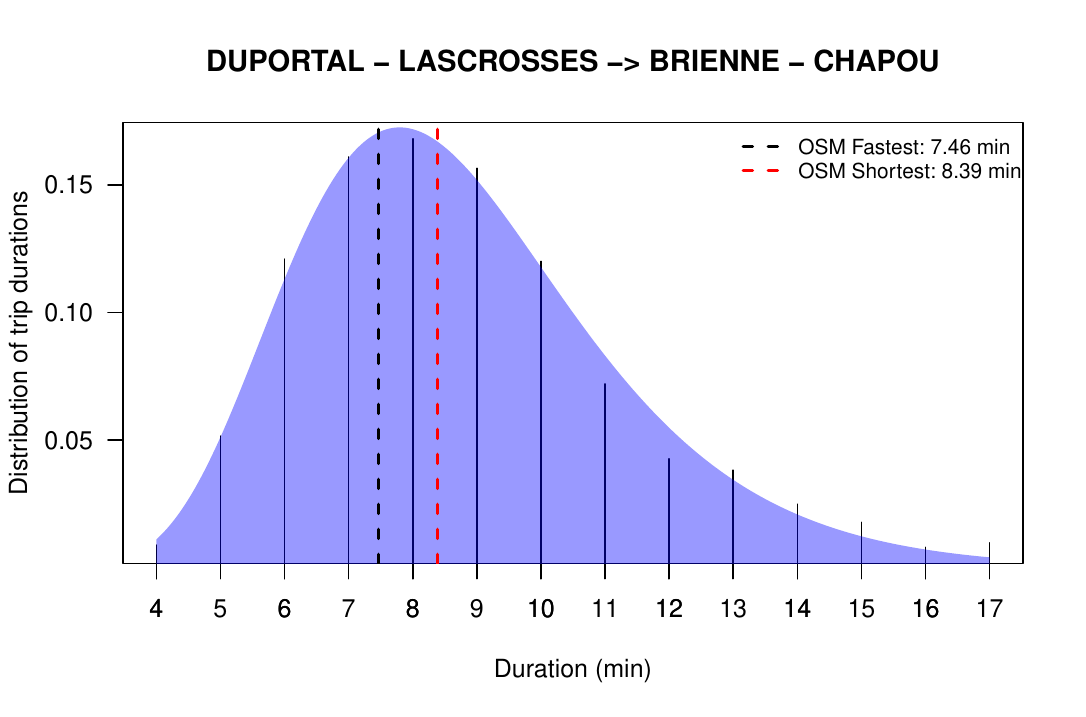} \hspace{1cm} \includegraphics[width = 0.45\textwidth]{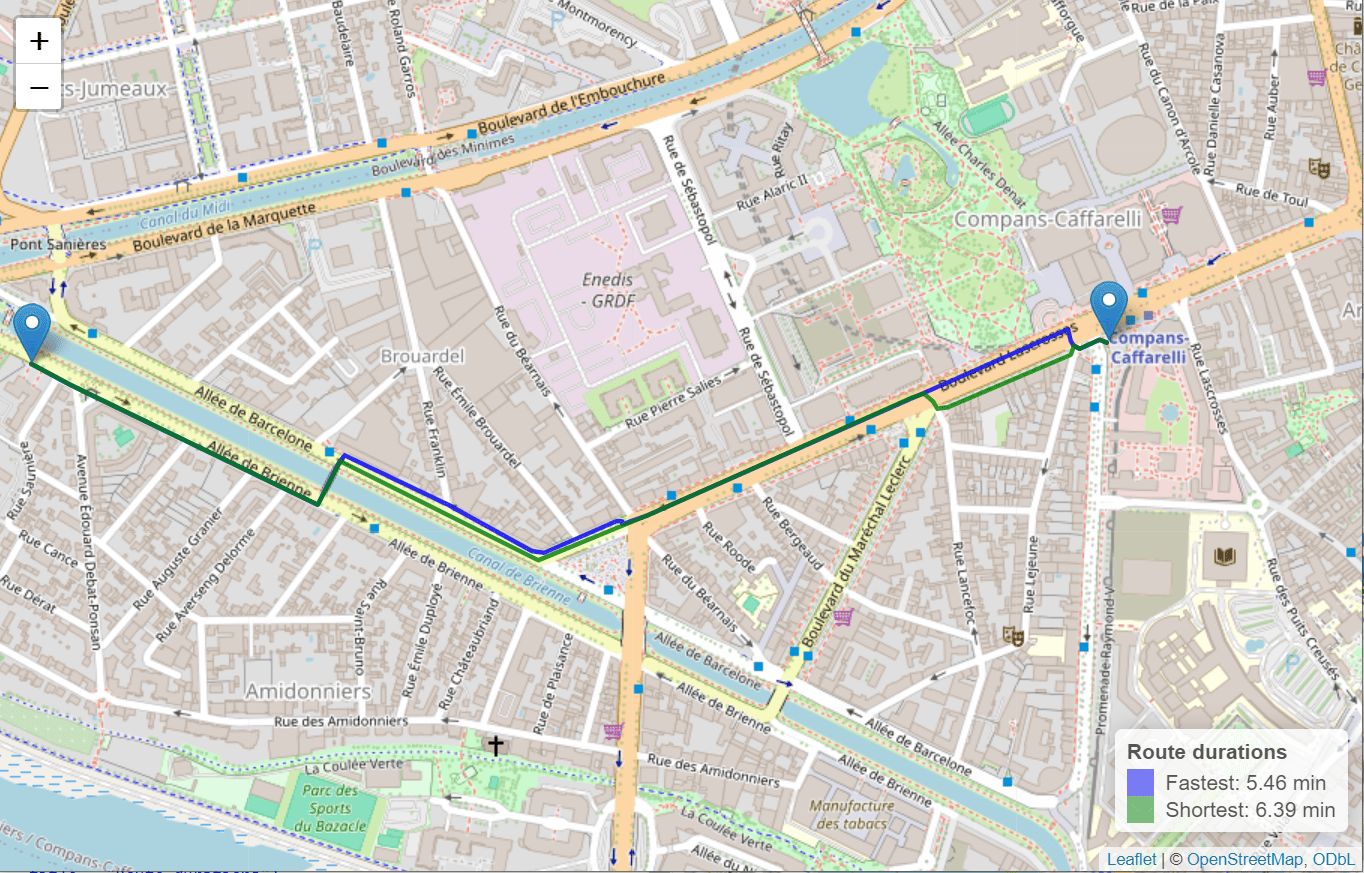} 
\includegraphics[scale = 0.7, width = 0.45\textwidth]{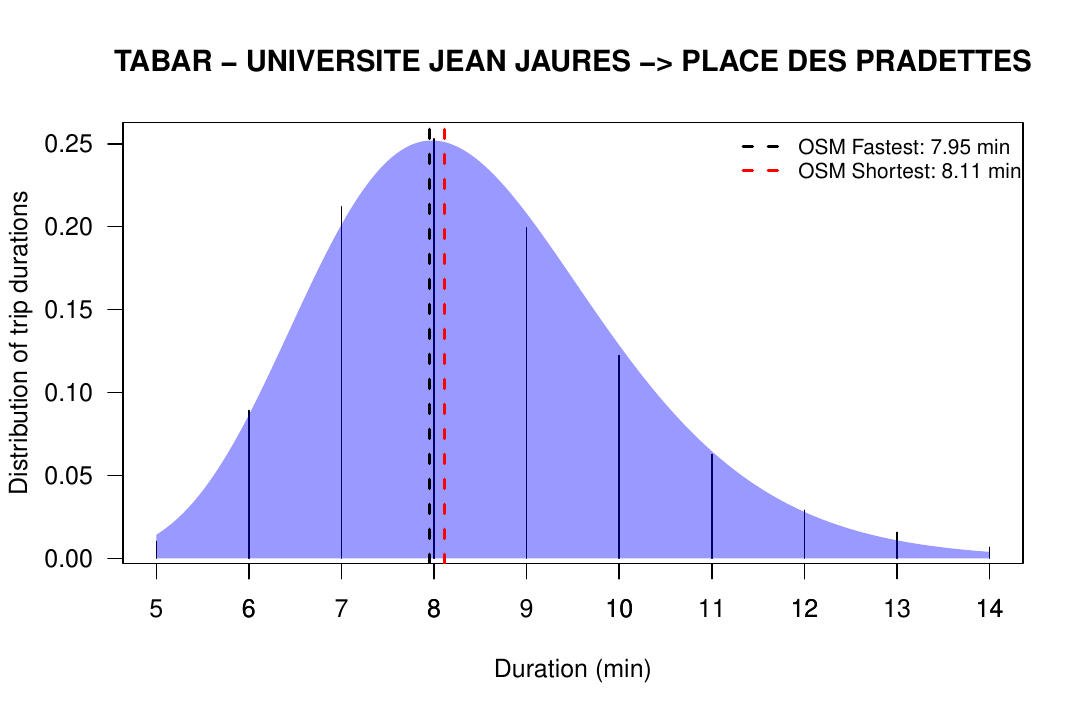} \hspace{1cm} \includegraphics[width = 0.45\textwidth]{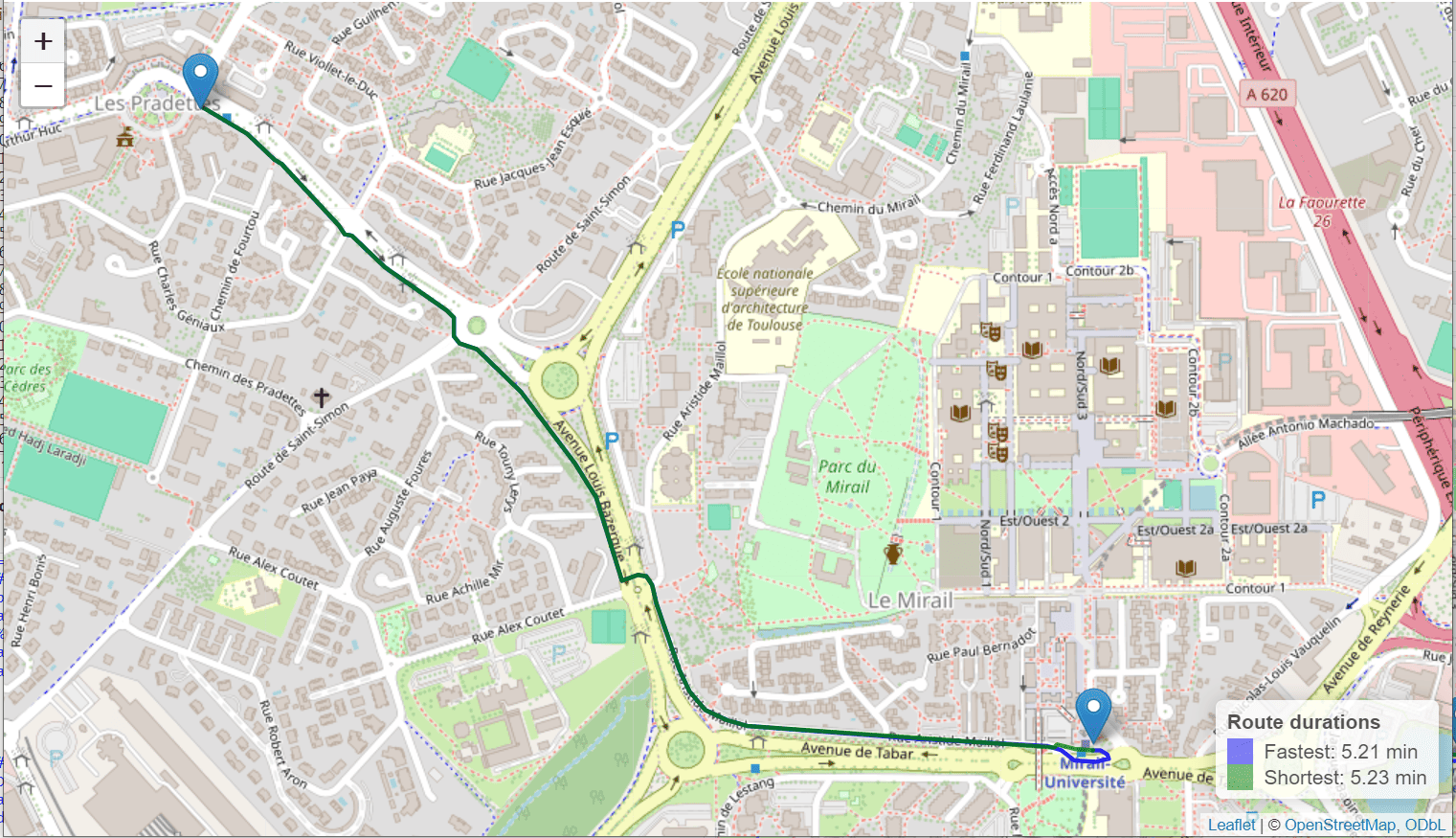} 
\end{center}
\caption{\footnotesize Examples of pairs of stations with only one route as confirmed by both OSM and the log-normal mixture model. }
\label{fig:un_group}
\end{figure}

 In $38.18 \%$ of cases, a single dominant behavior is detected in the sense that the hypothesis of a discretized log-normal model on the data is not rejected (to a level $\alpha =0.05$). In theses cases, we observe a strong relationship between the typical trip duration (mode of the log-normal distribution) and the duration of the main route provided by OSM, whether it is the shortest or fastest option. The scenario of a single dominant behavior is validated by OSM approximately $9$ times out of $10$, with the fastest and shortest routes being nearly identical. This situations is illustrated in the two examples of Figure \ref{fig:un_group}.

\paragraph{Diversity in BSS travel behaviors}~~\\

\begin{figure} [H]
    \centering
    \begin{subfigure}{0.5\textwidth}
        \centering
        \includegraphics[width=\textwidth,height=6cm]{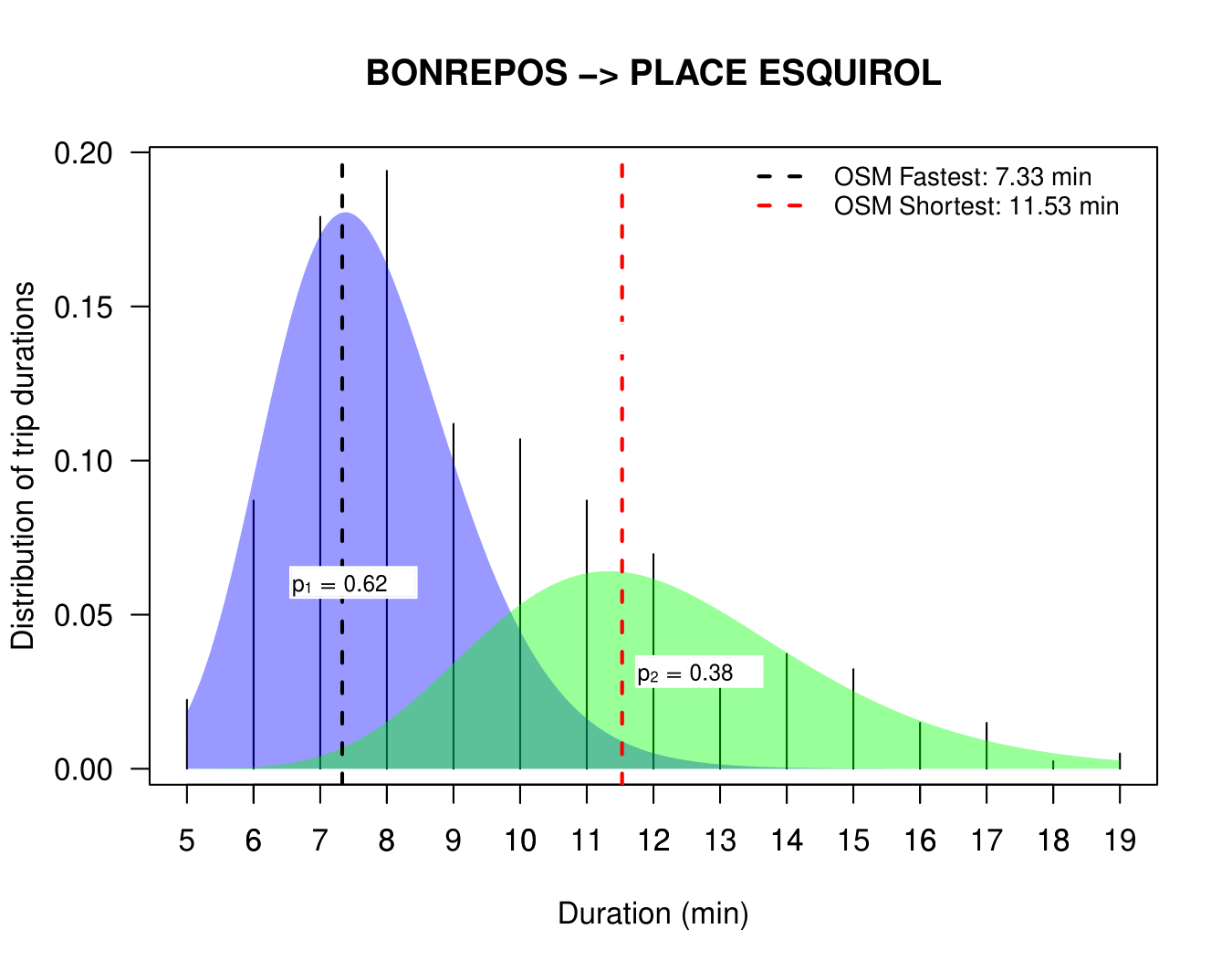}
    \end{subfigure}
    \hfill
    \begin{subfigure}{0.45\textwidth}
        \centering
        \includegraphics[width=\textwidth,height=5cm]{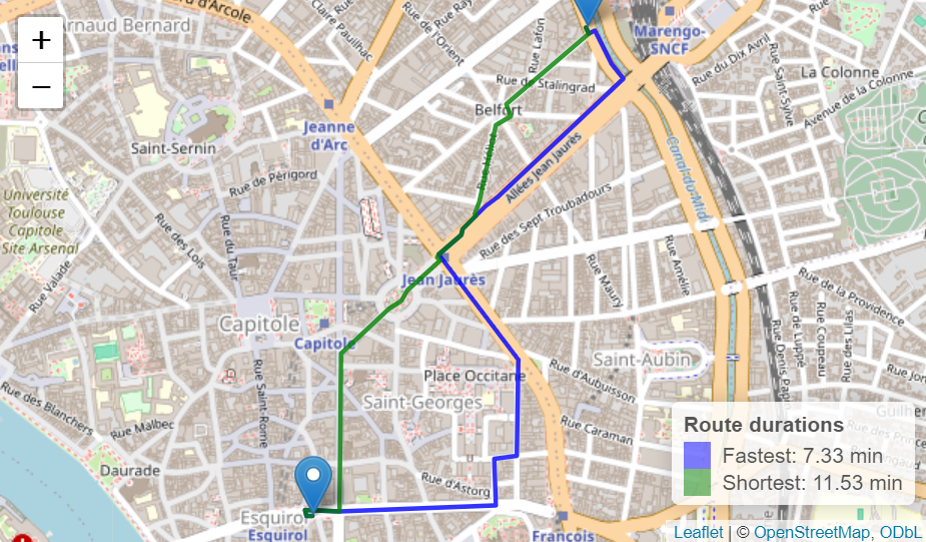}
    \end{subfigure}
    \caption{\footnotesize The figure on the left illustrates a mixture of log-normal distributions using the shortest and fastest durations provided by OpenStreetMap (OSM) for the route between Bonrepos and Place Esquirol. On the right, the map displays the two main routes suggested by OSM, with the typical route shown in blue, which corresponds to the first group, and the alternative route depicted in green, representing the second group of the mixture.}
     \label{fig:bonrepos_esq}
\end{figure}

In the remaining cases, accounting for $61.82 \%$ of the datasets, different cycling behaviors are detected and the use of a log-normal mixture model refines the analysis of cyclist practices by highlighting the possible coexistence of different travel behaviors within the same station pair. In these cases, the main component of the mixture remains strongly correlated with the duration of the fastest OSM route, as shown in Figure \ref{fig:bonrepos_esq}. Moreover, in $34.59\%$ of the cases where two components are identified in the mixture model, their modes closely match the fastest and shortest durations provided by OSM. This observation suggests that, in such situations, the second component captures a consistent alternative practice followed by a significant share of users. In the remaining $65.41\%$ of cases, the correspondence between the mixture components and the OSM reference routes is less clear. Several explanations may account for this. These cases may reflect heterogeneous practices that do not translate into clearly separated groups, but are instead aggregated within the same component. A single component may thus capture both optimized trips and others shaped by motivations such as comfort, safety, or routine, making interpretation less straightforward. In other instances, the modeling may not isolate OSM’s reference trajectories because of the closeness of travel times or due to inherent data variability (e.g., stops, detours, or uncertain navigation). Finally, interpreting the first component as systematically corresponding to the “main route” can be misleading, especially when several routes of similar duration coexist. These cases illustrate the limits of the mixture approach and emphasize the complexity of real-world cycling practices, which cannot always be cleanly aligned with theoretical routes suggested by routing algorithms.

This comparison with OSM provides an indication of how observed practices relate to the theoretical routes suggested by the platform. The probability associated with the main component of the mixture model can be interpreted as the share of users adopting the most direct or time-efficient behavior, while the secondary component reflects alternative or less regular practices. For instance, in the previous example, the behavior corresponding to OSM’s fastest route accounts for an estimated proportion of about $p \approx 0.62$ of users, according to the log-normal mixture adjustment.

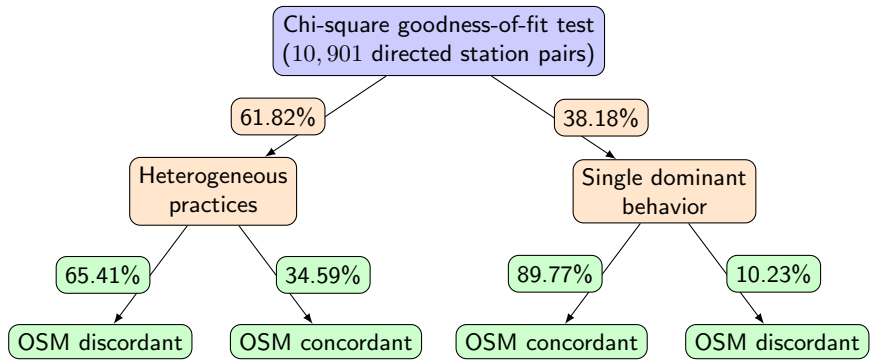
\begin{figure}[htbp]
  \centering
  \begin{tikzpicture}[
    grow=down,
    level 1/.style={sibling distance=60mm, level distance=20mm,
                    every node/.append style={fill=orange!20}},
    level 2/.style={sibling distance=30mm, level distance=20mm,
                    every node/.append style={fill=green!20}},
    edge from parent/.style={draw, -latex},
    every node/.style={font=\sffamily\small, align=center,
                       draw, rounded corners, fill=blue!20} 
  ]

  \node {Chi-square goodness-of-fit test\\ ($10,901$ directed station pairs)}
    child { node {Heterogeneous\\practices}
      child { node {OSM discordant}
        edge from parent node[left] {65.41\%} }
      child { node {OSM concordant}
        edge from parent node[right] {34.59\%} }
      edge from parent node[left] {61.82\%} }
    child { node {Single dominant\\behavior}
      child { node {OSM concordant}
        edge from parent node[left] {89.77\%} }
      child { node {OSM discordant}
        edge from parent node[right] {10.23\%} }
      edge from parent node[right] {38.18\%} };

  \end{tikzpicture}
  \caption{$\chi^2$ Test and OSM criteria}
  \label{fig:chi2_tree}
\end{figure}

\subsection{Quantifying cycling practices through mixture model modes, OSM routes, and cluster proportions}

\begin{figure}[H]
\begin{center}
\includegraphics[width = 0.45\textwidth]{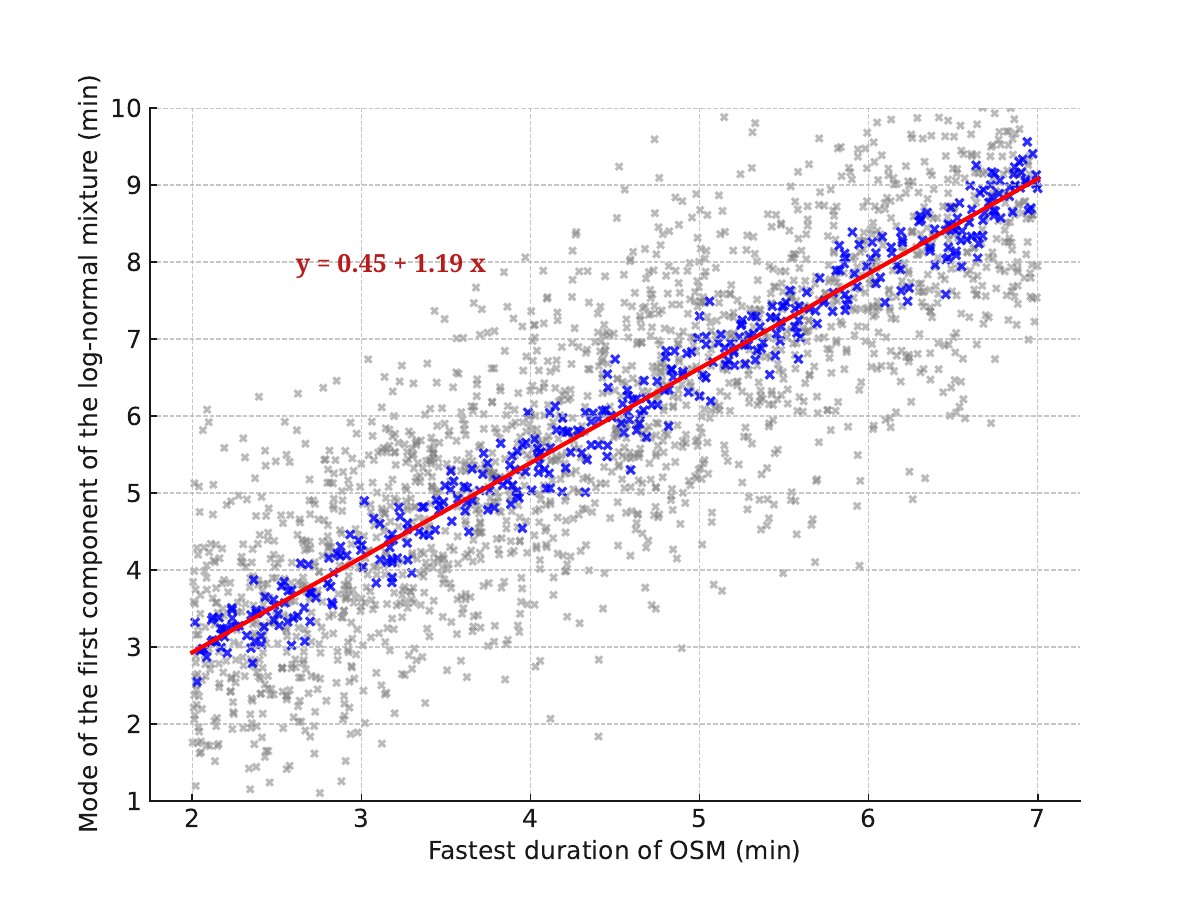} \hspace{1cm} \includegraphics[width = 0.47\textwidth]{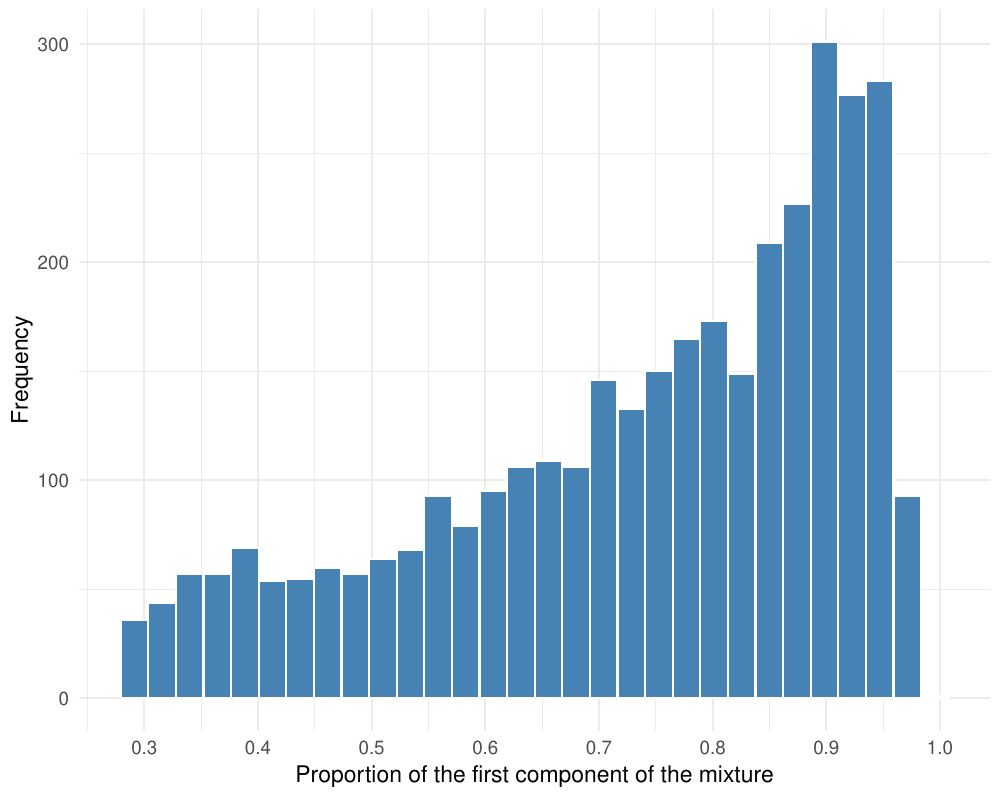} 
\end{center}
\caption{\footnotesize (left) Scatter-plot of the mode of the first component of the log-normal (mixture) model and OSMLX fastest route. Blue dots represent pairs of stations between which the data is well approximated by a single log-normal model, and gray dots those for which a mixture model is required. (right) Histogram of the first component of the mixture model for all station pairs.}  
\label{fig:nuage_points}
\end{figure}

Among the pairs for which a single dominant behavior is detected, a strong linear relationship appears between the typical trip duration in the log-normal model (defined by its mode) and the duration provided by OSM, as illustrated by the blue dots in the left plot of Figure \ref{fig:nuage_points}. Similarly, the mode of the first component of the log-normal mixture model also follows a comparable linear relationship with OSM’s fastest route (gray dots), although with greater variability.\\

Standard linear regression shows the clear linear relation between the mode of the first component in the (mixture) model and OSM's fastest route, 
($R^2  \approx 0.79$). This reveals that the true ``typical'' (or most likely) trip duration as observed from the data can be quite accurately predicted from the OSM's fastest route by applying an affine correction with slope $1.19$ and intercept $0.45$. The high accuracy of the linear model ($R^2 \approx 0.92$) for pairs of stations with a single route suggests that the two parameters may stem from real-life phenomena. The intercept of $0.45$, or about 27 seconds, could correspond to a fixed time loss induced with unlocking and/or locking the bike at the station's dock. The slope may indicate the loss of speed caused by using a less practical BSS bike compared to a regular one, the data suggesting here that a cyclist goes $1.19$ times slower with a BSS bike. It may also simply reflect the discrepancy between expected and observed speeds. The quality of fit significantly decreases when multiple clusters are identified, reflecting additional modeling error due to the difficulty of clearly distinguishing between alternative behaviors. This highlights the limitations of the log-normal mixture model in accurately capturing the complexity of route choice in such cases. \\


The mixture model makes it possible to quantify the proportion of riders who follow a dominant travel pattern versus those whose trajectories reflect more heterogeneous behaviors (see right panel of Figure \ref{fig:nuage_points}). The first cluster generally corresponds to the typical trip, which in many cases aligns with the fastest, shortest, or most direct route between two stations. This group is often composed of regular users or commuters, who tend to follow stable and time-efficient itineraries with little variability. These users are less inclined to deviate from their preferred path, which explains the strong consistency observed in this cluster. By contrast, the second cluster does not correspond to a single alternative route but instead aggregates a diversity of behaviors, including longer trajectories, intermediate stops, detours, scenic choices, or safety-oriented preferences.\\

The histogram in Figure \ref{fig:nuage_points} illustrates the distribution of the proportions associated with the first component of the mixture across station pairs. This distribution of proportions exhibits an increasing shape, indicating that in most cases the shortest route accounts for the largest share of users. For the majority of station pairs, between $70\%$ and $95\%$ of riders follow a common path, generally corresponding to the most direct or fastest option. Nevertheless, variability remains significant, as proportions range from 0.3 to 0.95. This means that all situations are represented, from cases where users are relatively evenly distributed across several itineraries to cases where a single route concentrates almost all trips. 
Overall, this distribution confirms that bike-sharing usage is largely structured by stable and homogeneous practices, while still revealing, in specific contexts, a plurality of routes that demonstrates the flexibility of urban cycling practices.\\

The proportion of users in the first group thus becomes a meaningful indicator of the prevalence of a stable and routine-like practice: the higher this proportion, the more strongly cyclists appear to converge toward a shared perception of efficiency or convenience. Conversely, a lower proportion suggests greater behavioral diversity, which may result from voluntary detours, occasional stops, safety considerations, or constraints linked to the local cycling environment. This ability to distinguish between dominant and heterogeneous practices represents a significant methodological advance, going beyond the descriptive capacities of conventional mapping or navigation tools. It also offers valuable insights into BSS usage dynamics at the local scale, by revealing how cyclists actually use the network and by informing mobility policies that better align with real-world practices.\\

These results give rise, on a case-by-case basis, to different interpretations : a clear distinction between two behaviors, the revelation of practices not identified by OSM, or the confirmation of a dominant trajectory with contextual variability. In the following section, we present a detailed analysis of selected station pairs that are representative of the different situations observed in the data.

\section{Detailed analysis of particular routes} \label{sec:particular}

In this study, an itinerary can be interpreted as the spatial materialization of a collective behavior. The mixture model does not merely identify theoretical paths (such as those proposed by OSM), but rather highlights regularities in cyclists’ practices. When a dominant behavior emerges, it often corresponds to the shortest or fastest route. Conversely, more heterogeneous behaviors may encompass several itineraries with similar durations, or reflect choices motivated by other factors such as stops, comfort, safety, or personal habits. Thus, referring to “behaviors” broadens the interpretation beyond the mere geometry of the route, while still maintaining a possible correspondence with the geographical reality of the trips.\\

To provide a concrete illustration of the diversity of configurations highlighted by our approach, we present below five case studies that are representative of the main types of situations encountered in the data. These range from trips where two distinct behaviors correspond to clearly separable itineraries, to cases where the statistical model reveals alternative practices not captured by OSM, as well as intermediate situations where a dominant route coexists with contextual variability, or cases where no significant alternative practice can be detected.

\subsection*{Clearly distinguishable behaviors along a dual itinerary}

Between  ENAC  and  Métro Rangueil,  a  total  of  194  trips  were  recorded  by  users  of  Toulouse’s bike-sharing system in 2022, of which 14 were identified as outliers and excluded from the analysis to ensure the quality of the statistical model fit. The adjustment using a log-normal mixture model shows excellent alignment between the estimated density and the observed empirical distribution, which highlights the relevance and robustness of the selected model.\\

\begin{figure} [H]
    \centering
    \begin{subfigure}{0.5\textwidth}
        \centering
        \includegraphics[width=\textwidth,height=6cm]{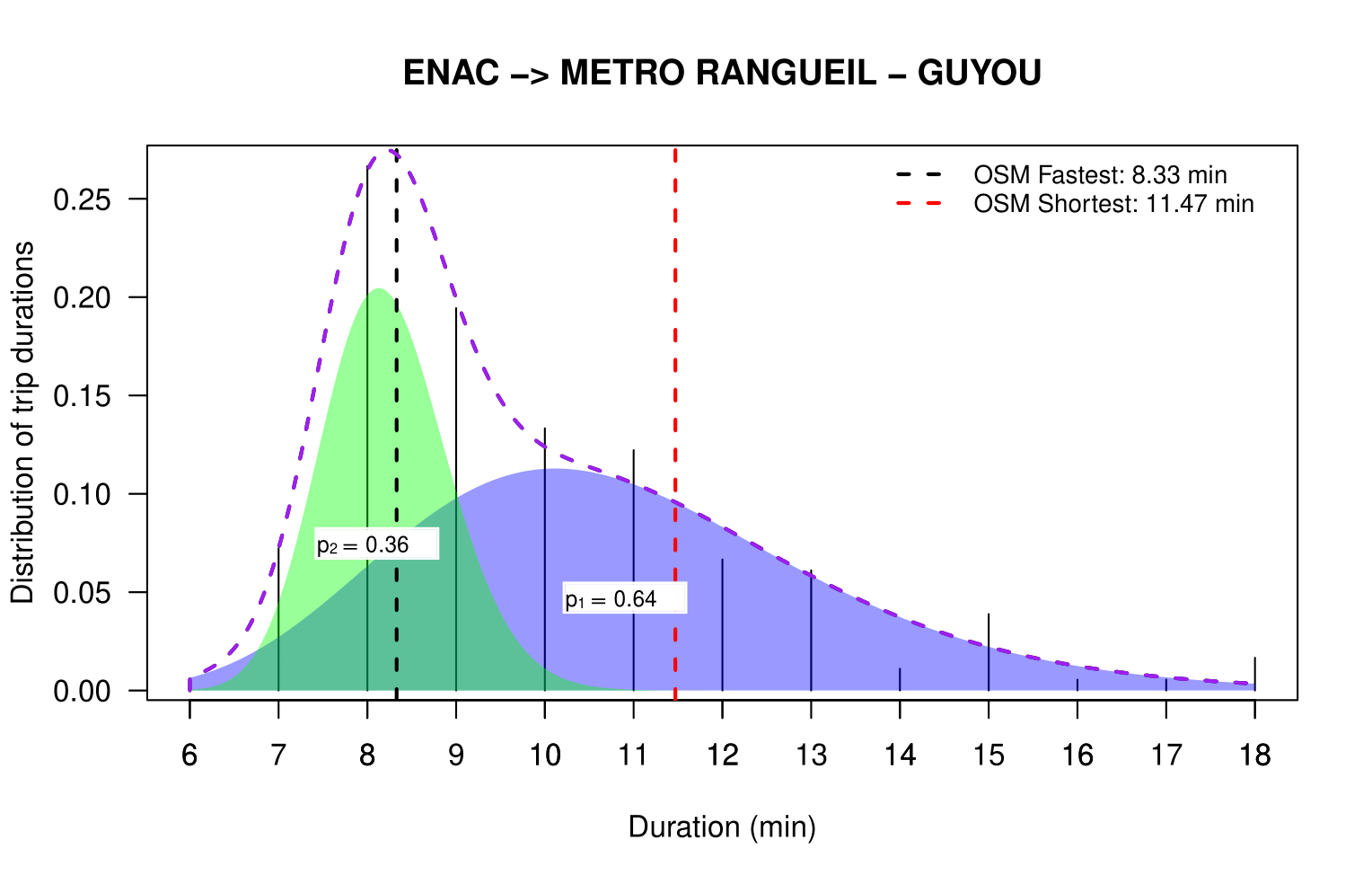}
    \end{subfigure}
    \hfill
    \begin{subfigure}{0.45\textwidth}
        \centering
        \includegraphics[width=\textwidth,height=5cm]{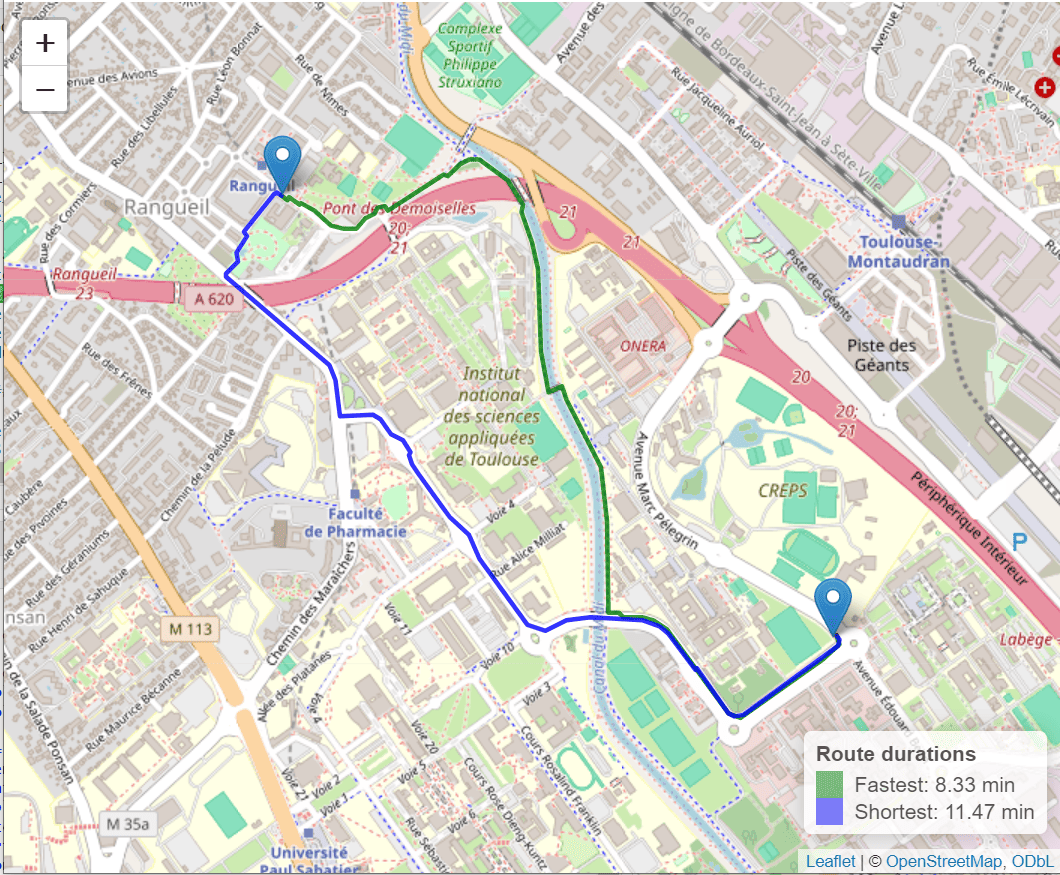}
    \end{subfigure}
    \caption{\footnotesize The figure on the left illustrates a mixture of log-normal distributions using the shortest and fastest durations provided by OpenStreetMap (OSM) for ENAC and Métro Rangueil. On the right, the map displays the two main routes suggested by OSM, with the typical route shown in blue, which corresponds to the first group, and the alternative route depicted in green, representing the second group of the mixture.}
    \label{specif2}
\end{figure}

The observed distribution exhibits an inverted structure to the case in Figure \ref{fig:bonrepos_esq}: the main component, which is relatively broad, is centered around 9.5 minutes (marked in blue in the figure \ref{specif2}), while a secondary, narrower component peaks around 7.5 minutes. This bimodal structure can be interpreted, through the mixture model, as the superposition of two distinct mobility behaviors. On the one hand, longer but consistent trips (they represent  $64\%$ of users), likely made by students or university staff who follow a standard route possibly more secure or comfortable even if slightly longer in duration. On the other hand, shorter trips, which may result from occasional shortcuts, favorable traffic conditions, or individual travel habits of cyclists familiar with the area. \\

This case is particularly interesting because the theoretical durations provided by OSM support this interpretation: the shortest route corresponds to the first (longer) component, while the fastest route aligns with the second (shorter) component. This reversal compared to typical situations where the shortest route is also the fastest suggests that the most direct path in this case may be constrained by factors such as topography, traffic signage, or the accessibility of certain cycling paths. Thus, this segment illustrates the value of using mixture models to detect sub-populations of trips and gain a deeper understanding of actual route choices, beyond theoretical estimations. It also emphasizes the importance of cross-referencing empirical data with network characteristics to better identify real-world urban cycling practices.\\

This analysis of the ENAC – Métro Rangueil segment clearly demonstrates the contribution of mixture models in providing a nuanced interpretation of travel behavior. The coexistence of two distinct groups of trips, well differentiated in terms of duration but not detectable through traditional mean-based methods, reveals the structural richness of urban mobility data. This observation also highlights the limitations of theoretical estimations provided by tools such as OSM when they fail to consider real-world usage constraints or accessibility issues. To further refine this analysis, it is useful to compare these findings with other station pairs that exhibit either similar configurations or markedly different distributions.

\newpage

\subsection*{Detection of an uncaptured itinerary through observed practices }

In 2022, 1058 trips were recorded between the Métro Faculté de Pharmacie and Belin ONERA stations. After removing outliers, the analysis was conducted on a refined dataset of 999 trips. These trips were likely made predominantly by students and employees, who used bicycles to complete the last leg of their trip from the metro station to their educational institution or workplace, particularly the ONERA research center.\\

\begin{figure} [H]
    \centering
    \begin{subfigure}{0.55\textwidth}
        \centering
        \includegraphics[width=\textwidth,height=6cm]{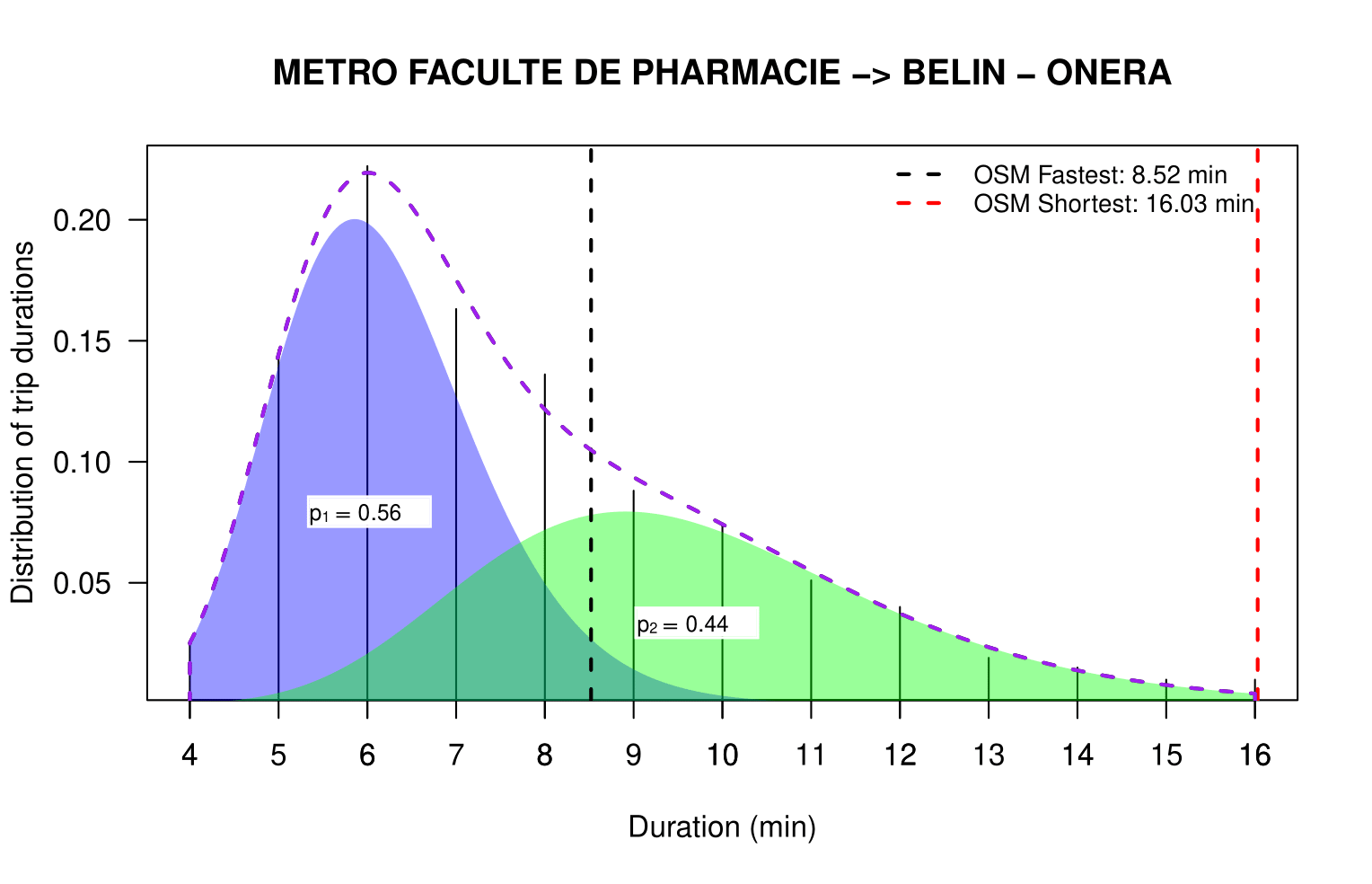}
    \end{subfigure}
    \hfill
    \begin{subfigure}{0.40\textwidth}
        \centering
        \includegraphics[width=\textwidth,height=5cm]{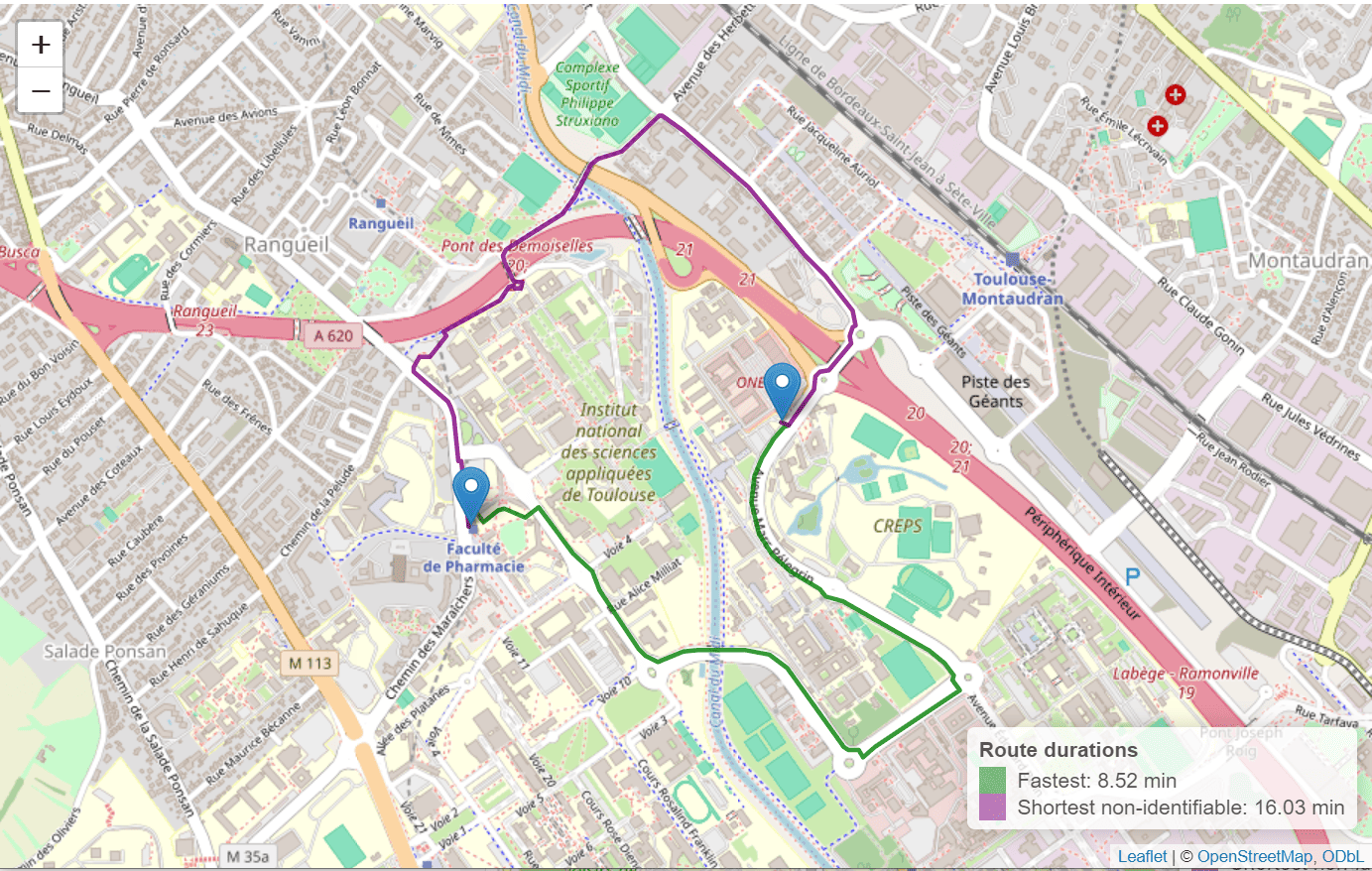}
    \end{subfigure}
    \caption{\footnotesize The figure on the left illustrates a mixture of log-normal distributions using the shortest and fastest durations provided by OSM for Métro Faculté de Pharmacie and ONERA. On the right, the map displays the two main routes suggested by OSM, with the typical route shown in blue, which corresponds to the first group, and the alternative route depicted in green, representing the second group of the mixture.}
    \label{specif1}
\end{figure}

The fitting of a log-normal mixture model for this route revealed the probable existence of alternative itineraries, a hypothesis already supported by the rejection of the null hypothesis in the goodness-of-fit test. The observed distribution displays two main components (gives Figure \ref{specif1}): a clear first mode around 6 minutes, and a broader second mode between 9 and 11 minutes. This case highlights the usefulness of mixture models: the first group likely corresponds to fast and direct trips, while the second group represents longer trips, potentially influenced by different urban conditions or route choices.\\

Interestingly, the fastest duration estimated by OSM aligns with the second component, which contrasts with the general trend observed in other cases. As for the shortest route suggested by OSM, it is not reflected in our empirical data. This suggests that OSM fails to identify the route associated with the first component, which is actually used by approximately $56\%$ of users on this segment. However, this route is clearly identifiable upon visual inspection of the map: it runs through the INSA campus, follows the Canal du Midi, and accesses the ONERA gate, which is restricted to students and staff, before reaching the Belin ONERA station.

\newpage

\subsection*{A dominant itinerary shaped by contextual variability }

The Sainte-Ursule → École Bonnefoy route provides a clear illustration of a case where one main itinerary dominates, while contextual variability introduces some dispersion in travel times. The empirical distribution is strongly concentrated around 8–9 minutes, which aligns well with the durations estimated by OSM for both the fastest and shortest paths. This suggests that most cyclists converge toward a common practice. Interestingly, the OSM reference durations fall slightly below the mode of the first component but match the regression line (see Figure \ref{fig:nuage_points}).\\

The presence of longer trips, extending up to 16 minutes, nevertheless indicates that occasional disruptions affect travel conditions. The fitted mixture model confirms this by identifying a secondary component, albeit small ($15\%$), which likely corresponds to contextual delays (traffic lights, temporary congestion, weather) or individual adjustments (short pauses, slight detours). Unlike other cases where two distinct peaks emerge, the observed asymmetry here indicates that these deviations do not represent a structurally different route but rather variations in how the main route is experienced.\\

This example highlights the ability of trip duration data to capture not only dominant route choices but also situational factors that introduce variability into otherwise homogeneous practices. It also illustrates a common scenario in dense urban environments: despite a moderately spread distribution, travel times are mainly explained by fluctuations around a single route rather than by the existence of fundamentally different itineraries.

\begin{figure} [H]
    \centering
    \begin{subfigure}{0.5\textwidth}
        \centering
        \includegraphics[width=\textwidth,height=6cm]{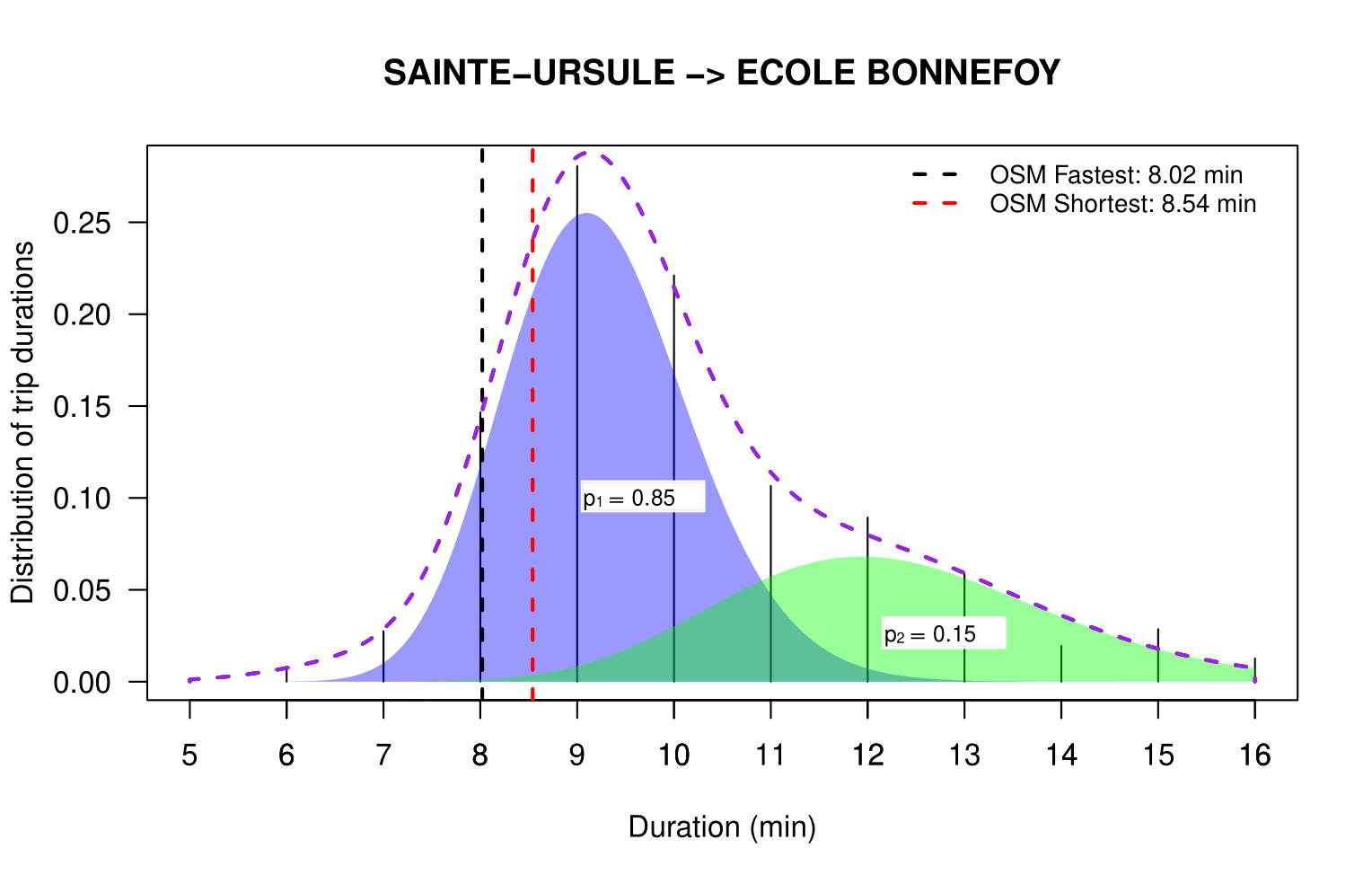}
    \end{subfigure}
    \hfill
    \begin{subfigure}{0.45\textwidth}
        \centering
        \includegraphics[width=\textwidth,height=5cm]{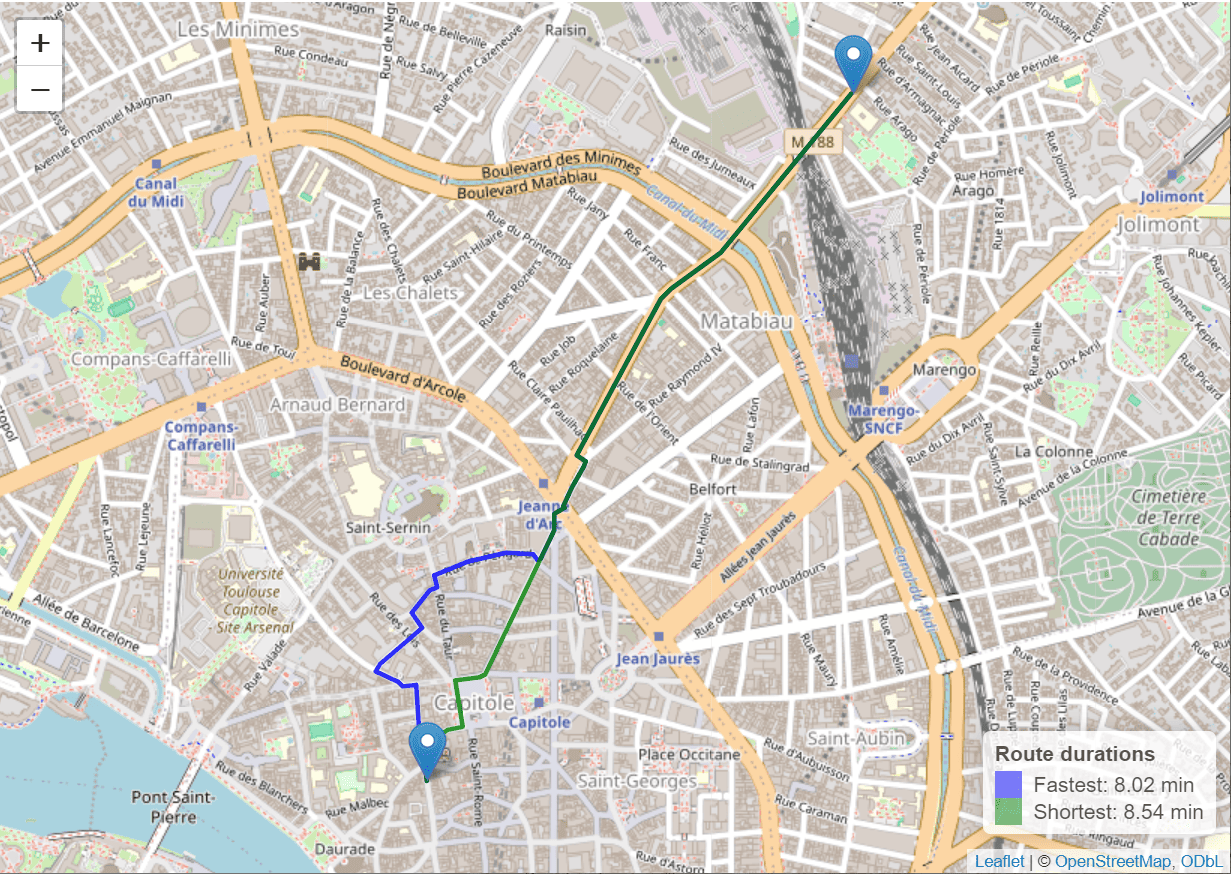}
    \end{subfigure}
    \caption{ \footnotesize
    The figure on the left illustrates a mixture of log-normal distributions using the shortest and fastest durations provided by OpenStreetMap (OSM) for Sainte-Ursule to Ecole Bonnefoy. On the right, the map displays the two main routes suggested by OSM, with the typical route shown in blue, which corresponds to the first group, and the alternative route depicted in green, representing the second group of the mixture.}
    \label{specif31}
\end{figure}

A similar scenario is observed for the Place Saint-Pierre → Place du Salin route, where OSM detects no alternative and proposes only one path. Yet, the fitted log-normal mixture model (validated by the initial goodness-of-fit test) reveals two distinct components, with about $57\%$ of users following the typical route. The comparison between the two cases is instructive: in the first, where $85\%$ of cyclists take the dominant route, they are mostly regular users, as tourists have little reason to go to École Bonnefoy, which is not a tourist area. In the second case, however, the lower proportion ($57\%$) of cyclists following the main itinerary suggests that the presence of tourists or occasional riders generates greater variability. The secondary component does not correspond to a truly alternative route but is more likely related to differences in speed or travel style, resulting in a bimodal distribution.

\begin{figure} [H]
    \centering
    \begin{subfigure}{0.5\textwidth}
        \centering
        \includegraphics[width=\textwidth,height=6cm]{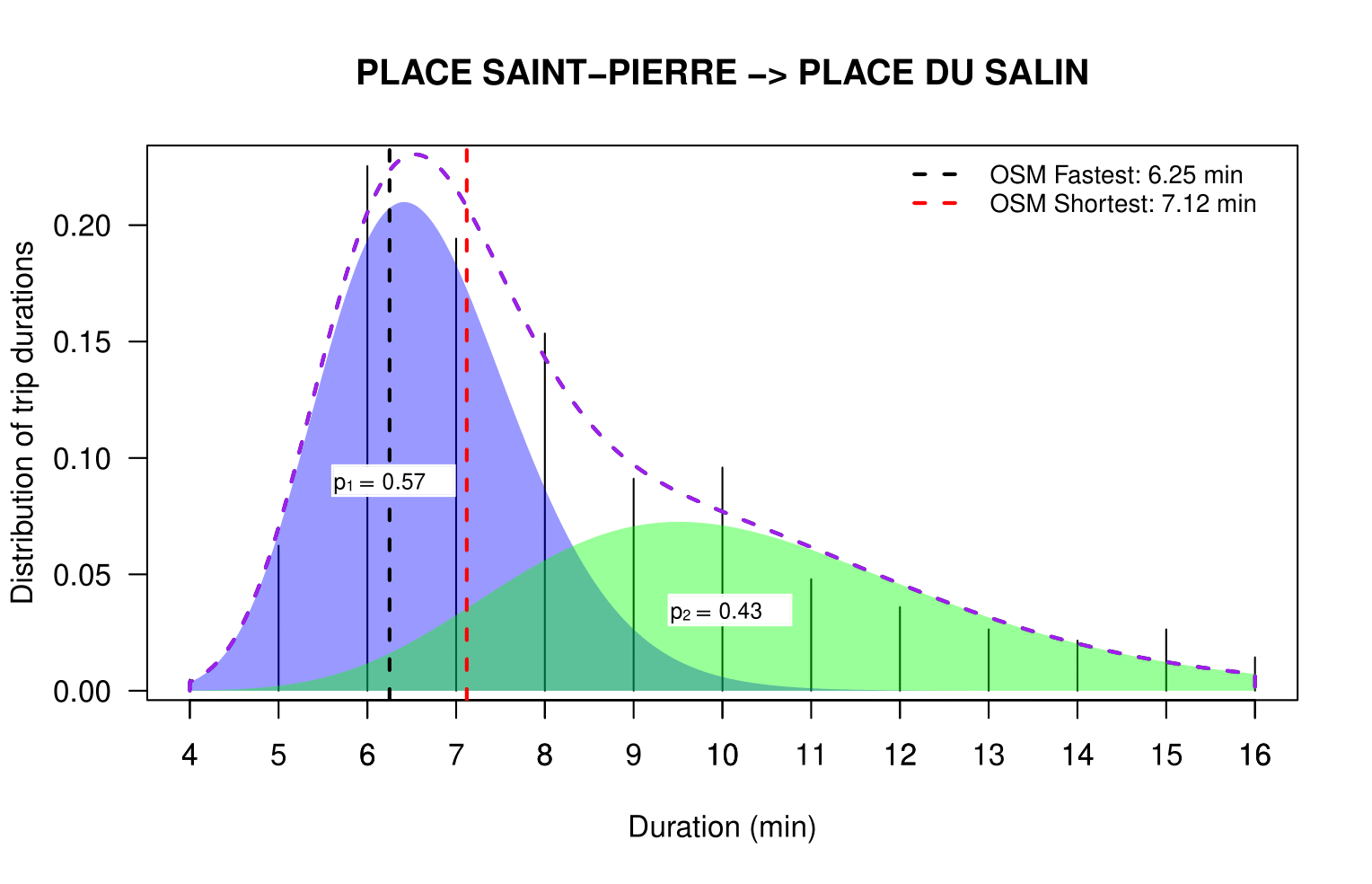}
    \end{subfigure}
    \hfill
    \begin{subfigure}{0.45\textwidth}
        \centering
        \includegraphics[width=\textwidth,height=5cm]{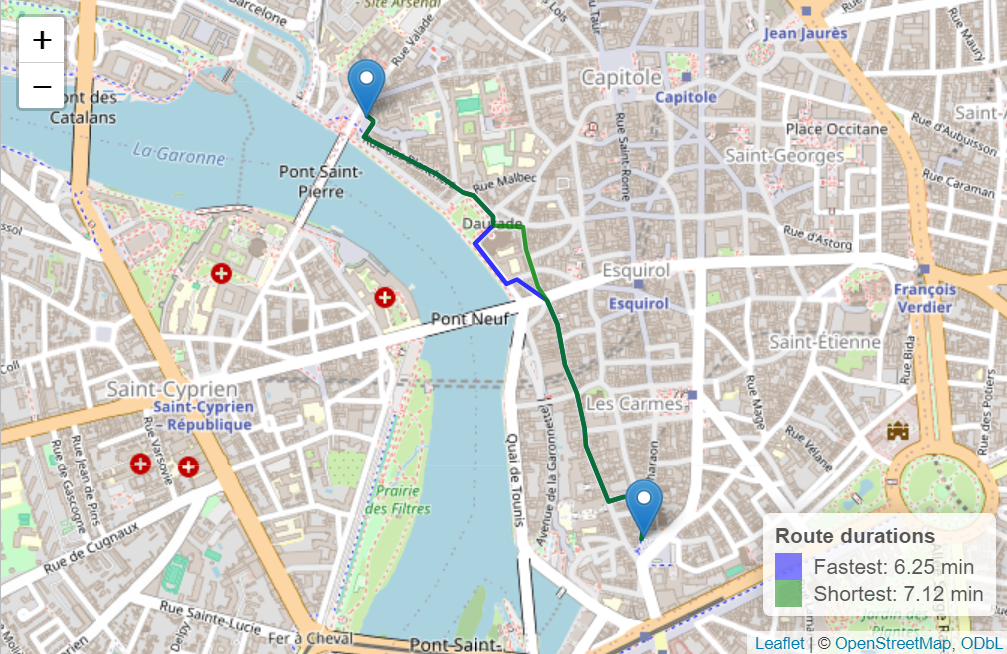}
    \end{subfigure}
    \caption{\footnotesize The figure on the left illustrates a mixture of log-normal distributions using the shortest and fastest durations provided by OpenStreetMap (OSM) for Place Saint-Pierre and Place du Salin. On the right, the map displays the two main routes suggested by OSM, with the typical route shown in blue, which corresponds to the first group, and the alternative route depicted in green, representing the second group of the mixture.}
    \label{specif3}
\end{figure}
    
\subsection*{A single itinerary without evidence of alternative practices} 

The route connecting Wagner–Brunhes to Place Saint-Pierre does not, according to the fitted model and the goodness-of-fit test to the log-normal distribution, provide clear evidence of alternative itineraries. The observed travel time distribution points to a single dominant mode, with no strong signs of multiple modes that would suggest distinct route choices.\\

Yet, OSM does identify two structurally different paths, with a marked difference between the fastest route (7.86 minutes) and the shortest (10.54 minutes). This discrepancy highlights a limitation of the modeling approach: empirical travel time data may not fully capture the complexity of the underlying network. Several explanations are plausible. Cyclists may consistently choose the faster route, leaving the slower one underrepresented in the sample. Alternatively, variability in travel times (due to traffic signals, congestion, or individual cycling speeds) may blur the distinction between different travel patterns.\\

In this sense, the gap between 7 and 8 minutes makes the fit particularly challenging. One could, by eye, attempt to draw two components, but these do not emerge clearly from the automatic estimation, underlining an inherent limitation of the statistical model.

\begin{figure} [H]
    \centering
    \begin{subfigure}{0.55\textwidth}
        \centering
        \includegraphics[width=\textwidth,height=6cm]{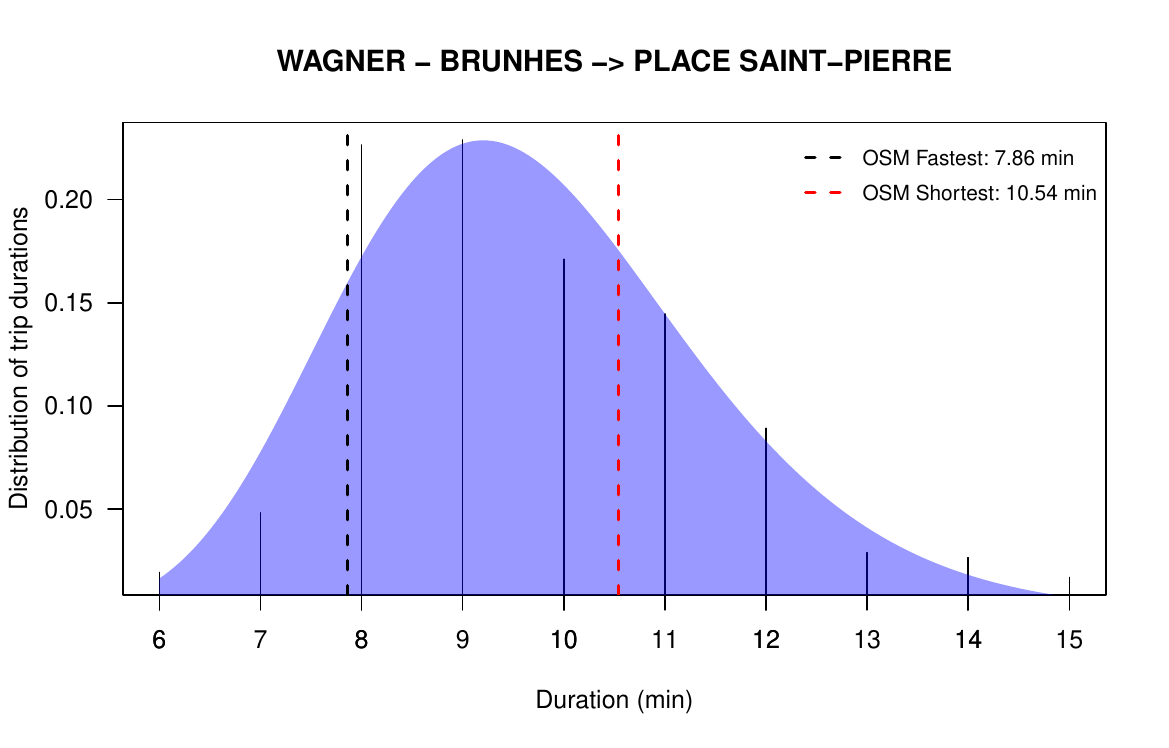}
    \end{subfigure}
    \hfill
    \begin{subfigure}{0.40\textwidth}
        \centering
        \includegraphics[width=\textwidth,height=5cm]{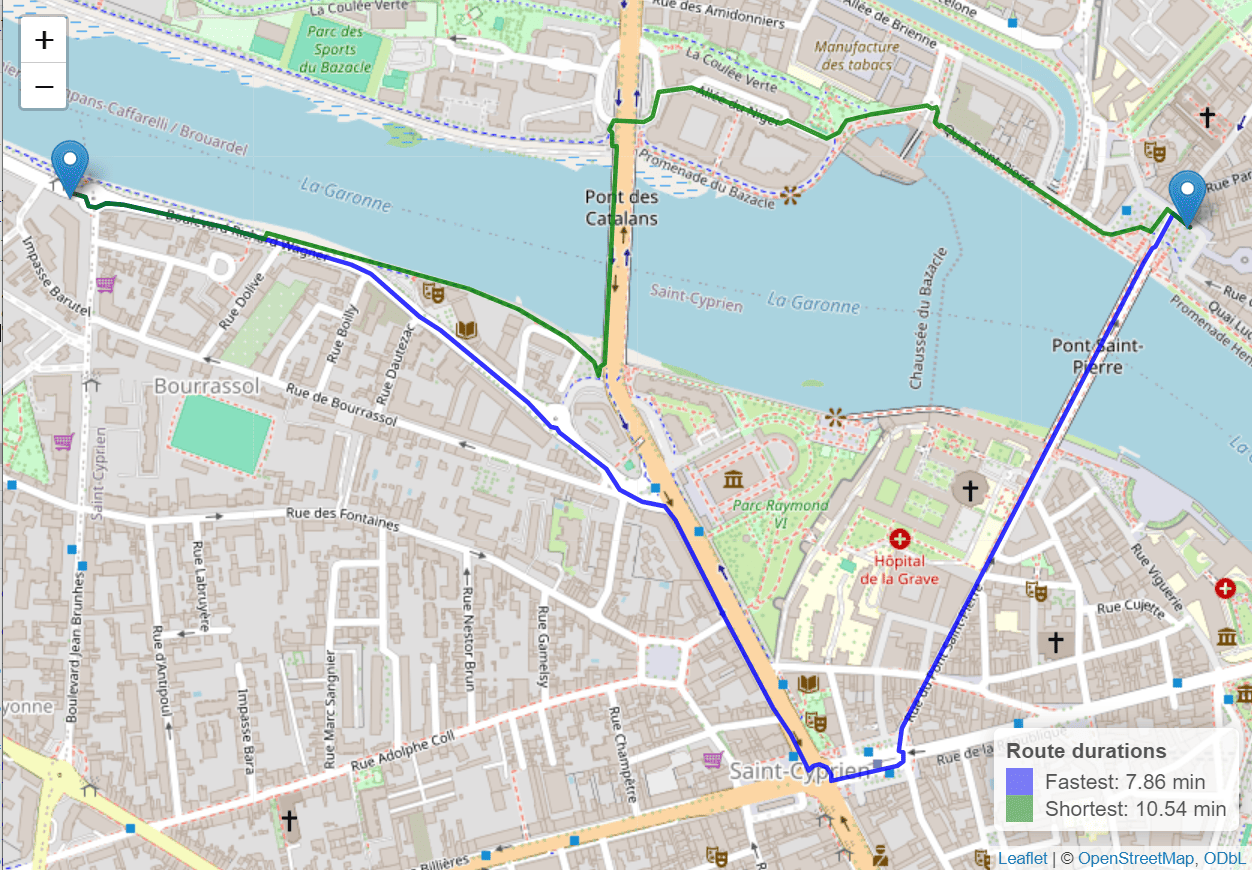}
    \end{subfigure}
    \caption{\footnotesize The figure on the left illustrates a non-mixture log-normal distribution fit for the trips durations in contrast <ith the shortest and fastest durations provided by OSM for Wagner-Brunhes to Place Saint-Pierre. On the right, the map displays the two main routes suggested by OSM.}
    \label{specif4}
\end{figure}

\section{Conclusion} \label{conclu}

This work proposes a statistical framework for analyzing the use of bike-sharing systems based on trip duration distributions, an approach that is particularly well suited to systems without GPS tracking. The use of log-normal mixture models makes it possible to identify both dominant cycling practices often associated with the fastest or most direct routes and more heterogeneous behaviors that include detours, intermediate stops, or alternative itineraries. This distinction allows us to move beyond the purely theoretical perspective of routing algorithms such as OSM, offering instead an empirical account of how cyclists actually appropriate the network.

Our results show that, for the majority of station pairs in Toulouse, a large proportion of users converge toward a dominant practice (the main component in two-component cases, or a single component when sufficient), reflecting the existence of routine or habitual trips. At the same time, a non-negligible share of trips corresponds to heterogeneous behaviors, underlining the importance of accounting for diverse motivations and constraints in urban cycling practices. These findings highlight the capacity of the statistical approach to confirm the prevalence of dominant routes while also revealing “hidden” practices that mapping tools alone fail to identify. Our methodological contribution thus provides a means of quantifying the coexistence of homogeneous and heterogeneous practices across an entire BSS network. Lastly, this approach could enhance and help improve digital maps (OSM, Google Maps, etc...) by updating or adding to current route calculation services which do not always fully reflect cyclists' real practices. \\

Despite promising results, our approach presents several limitations that must be acknowledged. In Toulouse, although the log-normal mixture model allows us to detect distinct behavioral patterns in over $60\%$ of cases, the detailed interpretation of the components can be challenging. In particular, the second component identified does not always correspond to a clearly defined alternative route but may instead encompass a variety of behaviors (such as detours, pauses, slowdowns, or context-driven choices) that are difficult to disentangle precisely. Moreover, when multiple routes have similar durations, the first component, which is assumed to represent the main route, may become ambiguous, thus reducing the robustness of the analysis. These limitations are further exacerbated in situations where the routes proposed by OSM do not match the ones actually used by cyclists, revealing a gap between theoretical paths and real-world practices. \\

Beyond its methodological scope, the proposed approach also carries practical implications. By highlighting discrepancies between theoretical itineraries and observed practices, it offers useful insights for urban planning and for the design of mobility policies better adapted to the real needs of users. Section \ref{sec:particular} illustrates these results through detailed case studies, ranging from station pairs dominated by a single route to those where mixture models reveal alternative practices not captured by OSM. Future extensions could integrate contextual variables (weather, infrastructure, user profiles), topographic information (distance), or continuous non-discretized data, in order to provide a more comprehensive picture of cycling behaviors.

\section*{Acknowledgements}
This research was supported by Région Occitanie (n° DOMSUB23003131) and Université de Toulouse (n°155 2024).
The authors would like to thank Maxime Lenormand, Angelo Furno and Bruno Revelli for their constructive comments and suggestions.


\bibliographystyle{alpha}
\bibliography{refs}

\end{document}